\documentclass[article]{WileyASNA-v1}
\usepackage[utf8]{inputenx}
\usepackage[T1]{fontenc}
\def\rfe{$R_\mathrm{FeII}$}
\def\feii{$\mathrm{Fe}${\sc ii}}
\def\feiiq{Fe{\sc II}$\lambda 4570$}
\def\hb{{\sc{H}}$\beta$\/}
\def\hbbc{{\sc{H}}$\beta_{\rm BC}$\/}
\def\kms{km\ s$^{-1}$}
\articletype{Review}%
\received{}
\revised{}
\accepted{}
\begin{document}
\title{The Main Sequence of quasars: the taming of the extremes\protect\thanks{Invited talk presented at the X{\sc iii} Serbian Meeting on spectral line shapes in astrophysics (SCSLSA)}}
\author[1]{P. Marziani*}
\author[2]{E. Bon}
\author[2]{N. Bon}
\author[3]{M. D'Onofrio}
\author[4]{B. Punsly}
\author[5]{M. \'Sniegowska}
\author[5,6]{B. Czerny}
\author[6,5]{S. Panda}
\author[6]{M. L. Mart\'{\i}nez Aldama}
\author[7]{A. del Olmo}
\author[7]{A. Deconto-Machado}
\author[8]{C. A. Negrete}
\author[8]{D. Dultzin}
\author[8]{T. Buendia}
\author[8]{K. Garnica}
\authormark{Marziani \textsc{et al}}
\address[1]{\orgdiv{Padua Astronomical Observatory}, \orgname{National Institute for Astrophysics (INAF)}, 
\orgaddress{\state{Padua}, 
\country{Country name}}}
\address[2]{ \orgname{Astronomical Observatory}, \orgaddress{\state{Belgrade}, 
\country{Serbia}}}
\address[3]{\orgdiv{Department of Physics and Astronomy ``Galileo Galilei''}, \orgname{University of Padua}, \orgaddress{\state{Padua}, \country{Italy}}}
\address[4]{\orgdiv{ICRA, Physics Department}, \orgname{University La Sapienza}, \orgaddress{\state{Rome}, \country{Italy}}}
\address[5]{\orgdiv{Nicolaus Copernicus Astronomical Center}, \orgname{Polish Academy of Sciences}, \orgaddress{\state{Warsaw}, \country{Poland}}}
\address[6]{\orgdiv{Center for Theoretical Physics}, \orgname{Polish Academy of Sciences}, \orgaddress{\state{Warsaw}, \country{Poland}}}
\address[7]{\orgdiv{Instituto de Astrof\'\i sica de Andaluc\'\i a}, \orgname{CSIC}, \orgaddress{\state{Granada}, \country{Spain}}}
\address[8]{\orgdiv{Instituto de Astronom\'\i a}, \orgname{UNAM}, \orgaddress{\state{Mexico}, \country{Mexico}}}

\corres{*Paola Marziani  \email{paola.marziani@inaf.it}}


\abstract{The last few years have seen the confirmation of several  trends associated with the quasar main sequence.  The idea of a main sequence for quasars is relatively recent, and  its full potential for the observational classification and contextualization of quasars’ properties has yet to be fully exploited. { The   main sequence drivers  are  discussed in terms of the properties of extreme  objects.}  We briefly summarize  developments that constrain the viewing angle of the    accretion disk in a particular class of quasars (extreme Population B, radiating at low Eddington ratio), as well as inferences on the chemical  composition of the broad line emitting gas, and on the nature of radio emission along the quasar main sequence. 
}




\maketitle

\footnotetext{\textbf{Abbreviations:} AGN: active galactic nuclei; BLR: broad line region: E1: Eigenvector 1; FWHM: full width half maximum; MS: main sequence;  NLSy1: narrow-line Seyfert 1;  Pop. A: Population A; Pop. B: Population B; RL: radio loud; RQ: radio quiet; SED: spectral energy distribution; xA: extreme Population A}

\section{Introduction: a main sequence   for type-1 (unobscured) quasars}\label{sec1}

\begin{figure}[t]
\centerline{\includegraphics[scale=0.45]{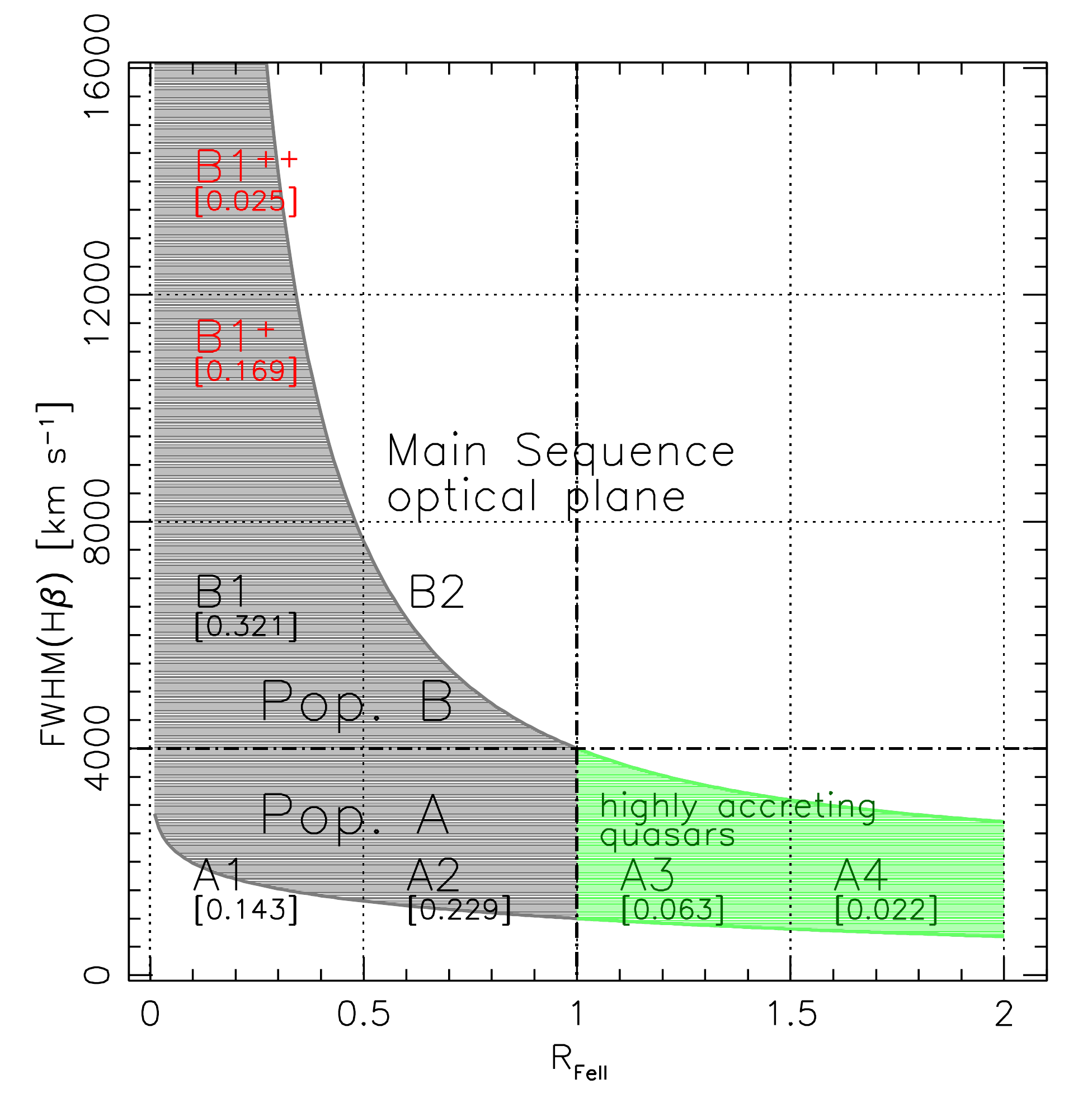}}
\caption{A sketch representing the occupation of MS quasars in the plane FWHM \hb\ vs \rfe, the MS “optical plane.” Each bin in the plane corresponds to a spectral type. The ones associated with extreme Pop.  B (red {labels}, top left) and the xA of highly-accreting quasars { at the high end of the \rfe\ distribution satisfying the condition \rfe $\ge 1$\ (green area) are the ones considered in this paper.   The numbers in square brackets below each spectral type designation are the source fractional occupation in an optically-selected sample \citep{marzianietal13a}. } \label{fig:ms}} \end{figure}

The main sequence (MS) concept originated from  a Principal Component Analysis  on the spectra of Palomar Green  quasars that allowed the identification of  a first Eigenvector  (Eigenvector 1, E1). The E1 was found to be associated with an anti-correlation between strength of the singly ionized iron emission blend centered at $\lambda$ 4570 (\feiiq) and the FWHM of  H$\beta$ \ (or peak intensity of [O{\sc iii}] 4959,5007). The MS itself can be effectively represented in a plane where the FWHM H$\beta$ data are diagrammed against the parameter \rfe\, defined as  the flux ratio between \feiiq\ and the broad component of \hb\ i.e., \rfe = F(\feiiq)/F(\hbbc). Figure \ref{fig:ms} shows a sketch of the occupation of the quasar MS in this parameter plane.

 The MS has withstood the test of time; from the original 80 sources in the paper by \cite{borosongreen92} and the tens of objects in an even earlier study by \citet{gaskell85}, the MS has been detected and studied in sample of $\sim$ 200 objects \citep{sulenticetal02}, $\sim$ 500 \citep{zamfiretal10}, to $\sim$ 20000 \citep{shenho14}. The main trends  have been confirmed, extended to multifrequency data  related to the accretion process and the accompanying outflows  (\citealt{sulenticetal00a,sulenticetal00b}, see also the summary Table of \citealt{fraix-burnetetal17}), and revealed also by sophisticated techniques such as locally linear embedding in manifold learning \citep{jankovetal21}.

Why are the  seemingly-obscure parameters FWHM \hb\ and \rfe\ so revealing? FeII emission is self-similar in type-1 AGN but the relative { flux} to \hb\ (parameterized by \rfe) can vary from undetectability to \rfe\ $\gtrsim 2$, with values larger than $\approx$ 2 being exceedingly rare. The \feii\ emission extends from UV to the IR and can dominate the thermal balance of the low-ionization broad-line region  \citep[BLR; ][]{marinelloetal16}. The FWHM of \hb\footnote{For the sake of simplicity we shall omit the subscript BC from now on.}\ is related  to the velocity field in the low-ionization part of the BLR  (predominantly  virialized) and to viewing angle effects. Therefore these two parameters reflect important aspects of the emitting gas physical conditions and dynamical status, along with the orientation (defined as the angle $\theta$\ between a putative axis of symmetry and the line of sight  at  which the AGN is seen). {The MS includes only type-1 AGN i.e., sources for which broad lines are visible in natural light. According to the unification schemes, $\theta$ is constrained between 0 and $\sim$ 45 — 60 degrees \citep[][see also \citealt{marinetal16} for a more recent perspective]{antonucci93,urrypadovani95}. Effects of changing $\theta$\ on the line width are believed to be significant for radio-quiet type-1 AGN as well, although there is not as yet an established view to connect $\theta$\ to observed spectral parameters.}  Orientation definitely plays a role in beamed radio emission \citep{urrypadovani95}, although the effects on the broad optical and UV lines are subject of current debate (more in \S \ \ref{orien}). 

The elbow-shaped MS allows for the definition of two main populations: Population A and B \citep[][{hereafter Pop. A and Pop. B respectively}]{sulenticetal00a}  on the basis of a  limit on the FWHM  of \hb\ ($\approx$ 4000 \kms\ at low and moderate luminosity),  as well as of several spectral types in narrow ranges of FWHM and \rfe. The Population A and B spectra are usually so very different that their classification can  be recognized by eye:  Pop. A sources show sharp \hb\ profiles, prominent \feii, and weak [O{\sc iii}]$\lambda\lambda$4959,5007 emission; on the converse a typical Pop. B spectrum shows Gaussian-line \hb\ profiles, weak \feii, and strong and spiky  [O{\sc iii}] (see the Figures in \citealt{sulenticetal00a}). If one compares a high-ionization line such as C{\sc iv}$\lambda$1549\ to \hb, the first shows a significant shift to the blue with respect to the quasar rest frame.\footnote{Most precisely traced by low-ionisation narrow emission lines such as [O{\sc ii}]$\lambda$3727 \citep{bonetal20}.} Summing up decades of works in monitoring and in the study of the spectroscopic properties of quasars,  we can say that the emitting region can be heuristically subdivided in two sub-regions: a low ionization, closely associated with the accretion disk, and that has been found to be predominantly in virial dynamical equilibrium, for both Pop. A and B (even with significant differences that are still poorly understood), and an outflow/wind region \citep[][]{collinsouffrinetal88,elvis00}, by far more evident in Pop. A and especially in extreme Population A (xA, defined by \rfe $>$ 1; see Fig. \ref{fig:ms}).  { The aim of this paper is to emphasize the {\em danger} of ignoring the trends of the main sequence, by showing how different the objects at the opposite extreme ends are. } 

Before discussing the interpretation of the main sequence { (Section \ref{interp})}, it is helpful to consider why there is a special reason to prefer  4000 \kms, instead of    2000 \kms\  for the \hb\ FWHM limit. Most studies still distinguish narrow-line Seyfert 1 (NLSy1s, defined from the condition FWHM \hb $\lesssim$ 2000 \kms) as a distinct class, and compare NLSy1s to “broad-line” AGN. However, the \hb\ profiles remain Lorentzian-like  up to  around FWHM(\hb)  = 4000 \kms, and the  C{\sc iv}$\lambda$1549 blueshifts remain consistent up to around the same FWHM  limit \citep{marzianietal18a}. A clear change in the \hb\ line profiles occurs at FWHM(\hb)  $\approx$ 4000 \kms: composite profiles in the range  3000 \kms\ $\lesssim$ FWHM(\hb)  $\lesssim$ 4000 \kms\ are best fit with a Lorentzian function, while for 4000 \kms\ $\lesssim$ FWHM(\hb)  $\lesssim$ 5000 \kms\ a double Gaussian provides a best fit \citep{sulenticetal02}.   

{The overview of the MS interpretation  (Section \ref{interp}) is followed by an analysis of  chemical composition and orientation effects among Pop. B sources (Section \ref{extremeb}). The results on chemical compositions are mirrored by the ones obtained for Pop. A (Section \ref{extremea}). In the case of xA, a novel result summarized in \S \ref{radio} is the connection between the defining property of strong \feii\ emission and significant radio power, that mirrors the ``jetted'' origin of the tremendous radio power of some 20 \% of Pop. B sources \citep{zamfiretal08}. Our conclusion (Section \ref{concl}) stresses how the improvement in our understanding of quasar physics might be instrumental in the use of xA quasars as cosmological probes. }

\section{Interpretation of the Main Sequence}
\label{interp} 

Several approaches consistently support a relation between $L/L_\mathrm{Edd}$\ and \rfe. For example,   the average  of the stellar velocity dispersion of the host galaxy in narrow luminosity bins (a proxy for $M_\mathrm{BH}$)   decreases with \rfe\ implying that $L/L_\mathrm{Edd}$\  increases with \rfe\ \citep{sunshen15}. The $L/L_\mathrm{Edd}$\ computed from the virial black hole mass $M_\mathrm{BH}$ relation for sources with reverberation mapping data is correlated with \rfe\  \citep{duetal16a}. 

A toy model that assumes a virial relation  for $M_\mathrm{BH}$ in the form:
\begin{equation}
M_\mathrm{BH} = f(\theta) \frac{r_\mathrm{BLR} \mathrm{FWHM}^2}{G}
\end{equation}
where $f(\theta$) is the viewing-angle dependent virial factor shows that the occupation  of the MS  optical plane at low-$z$  can be accounted for  in terms of  Eddington ratio and orientation \citep{marzianietal01}. 

\begin{figure}[t]
\centerline{\includegraphics[scale=0.45]{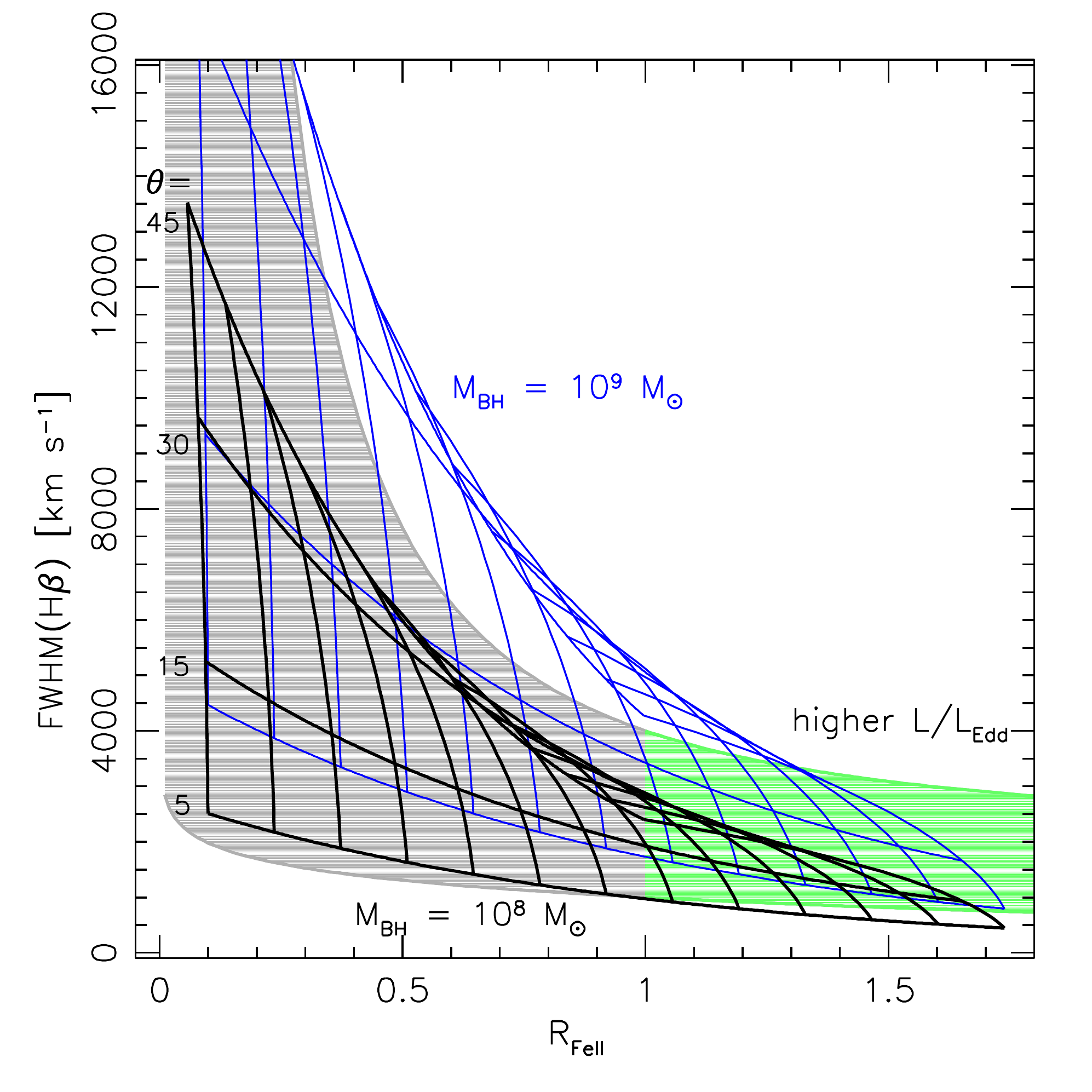}}
\caption{The occupation of MS quasars in the plane FWHM \hb\ vs \rfe\ as in Fig \ref{fig:ms} with two grids of $L/L_\mathrm{Edd}$\ and $\theta$, for $M_\mathrm{BH} = 10^8$ M$_\odot$\ (black) and $M_\mathrm{BH} = 10^9$ M$_\odot$\ (blue). Eddington ratio increase along with \rfe, while viewing angle $\theta$  decreases toward the bottom.   \label{fig:msgrid}} \end{figure}

{ Figure \ref{fig:msgrid} shows the prediction of the toy model overlaid to the shape representing the occupation of the AGN in the optical plane of the MS. A surprising aspect of the MS is that the sources radiating at higher $L/L_\mathrm{Edd}$\ are also the ones showing a spectrum of lower ionization degree. Here we just recall that the basis of the toy model (developed in \citealt{marzianietal01}, and expanded in  \citealt{pandaetal19}, where a systematic analysis is presented; see also \citealt{marzianietal18} for a summary description) is the connection between ionization parameter and \feii\ prominence that requires a systematic change in gas density and other physical parameters such as metallicity $Z$. Orientation effects are then compounded assuming a virial velocity field in a flattened configuration, and possible limb darkening effects on the \feii\ emission.} Not only $L/L_\mathrm{Edd}$\ is roughly proportional to \rfe, but also the elbow shape of the MS is accounted for by the combination of orientation and $L/L_\mathrm{Edd}$\ variation { (Fig. \ref{fig:msgrid})}.  Orientation yields an almost vertical (i.e., parallel to the FWHM axis) displacement from broader to narrower profiles, passing from viewing angle $\theta \approx 50$ to   $\theta \approx 5$. Population B sources (broader, with \rfe\ $\lesssim$ 0.5) are associated with low Eddington ratio ($L/L_\mathrm{Edd}\lesssim 0.25$), while Pop. A sources are mostly associated with higher Eddington ratio. { Extreme Pop. B means that Eddington ratio can be as low as $L/L_\mathrm{Edd}\lesssim 0.01$, with by definition  very broad, but also often double-peaked or irregular Balmer line profiles  \citep{gancietal19}. On the converse, xA means sources that are radiating at maximum luminosity-to-black hole mass, at high (possibly super-Eddington) accretion rates \citep[][see also \citealt{donofrioetal21} for a recent review]{marzianietal14}.\footnote{According to current accretion disk theory, the accretion rate can be arbitrarily high but the luminosity saturates to a limiting value with ($L/L_\mathrm{Edd}  \sim 1$ \citep{mineshigeetal00}. } }

An interesting inference from the grid shape is that some Pop. B might appear as Pop. A because of the effect of orientation, since the observed FWHM $\propto \delta v_\mathrm{K} \cdot \sin \theta$\ {where the $v_\mathrm{K}$ is the intrinsic Keplerian velocity field}. For $\theta \rightarrow 0$, the FWHM might become FWHM \hb $\lesssim$ 4000 \kms. These sources are however expected to be rare, since the probability of observing a source at angle $\theta$ is also $P(\theta) \propto \sin \theta$. 
Intrinsic variability may move the sources across the main sequence. As we progress towards the B1$^{++}$, the variability amplitude becomes larger, and   we may have   variations of FWHM as well for the same object and therefore same mass and inclination, so we have even larger areas on this plot that an  object could take. A case study is provided by the extensively-monitored Pop. B source NGC 5548 \citep{bonetal18a}: the   parameters FWHM \hb\ and \rfe\ constrain the source within the B1, with an occasional excursus in bin B1$^{++}$, in correspondence of the luminosity minimum. We are still waiting for a changing look AGN that transitions from a Pop. B to a Pop. A spectral type following a large increase in luminosity; this phenomenon  would provide further positive evidence of the spectral type association with Eddington ratio.\footnote{A change from Pop. B to Pop. A in response to a luminosity increase would disprove the current interpretation of the Eigenvector 1 MS, and is not expected.}

If the BLR radius follows {a scaling relation in the form of a} power-law with luminosity ($r \propto L^\mathrm{a}$, \citealt{kaspietal00,bentzetal13}), under the standard virial assumption, then {the line FWHM can be written as a function of the bolometric luminosity, black hole mass $M_\mathrm{BH}$, and $\theta$-dependent virial factor:}
 \begin{eqnarray}
\mathrm{FWHM} &  \propto & f_\mathrm{S}(\theta)^{-\frac{1}{2}} \left(\frac{L}{M_\mathrm{BH}}\right)^{-\frac{a}{2}} M_\mathrm{BH}^\frac{1-a}{2}\\
&  \propto & f_\mathrm{S}(\theta)^{-\frac{1}{2}}  {L}^\frac{1-a}{2} \left(\frac{L}{M_\mathrm{BH}}\right)^{-\frac{1}{2} }
\label{fwhm}.     
\end{eqnarray}

If  $a = 0.5$, one can easily see that FWHM $\propto  M_\mathrm{BH}^\frac{1}{4}$\ for   fixed $\theta$\ and $L/L_\mathrm{Edd}$,  implying that the grid is shifted upwards and that the FWHM increases by a factor $\sim 1.8$ for a ten-fold increase in $M_\mathrm{BH}$. Going one step further   it is possible to account for the \rfe\ values in each spectral bin assuming a systematic increase of density, chemical abundances and Eddington ratio from the low \rfe, high FWHM to the high \rfe, low FWHM end of the MS \citep{pandaetal19}, keeping a consistency with the  $r_\mathrm{BLR}$\ values expected from the scaling $r_\mathrm{BLR} - L$\ \citep{bentzetal13}, and including the correction derived for highly accreting quasars \citep{martinez-aldamaetal20}.

\section{Extreme Population B }
\label{extremeb}

The extreme Population B involves by definition sources with broad Balmer profiles {(a conventional limit could be set at $\approx 10000$ \kms\ of at 12000 \kms\ to include only the most extreme sources}, and very low (often undetectable) \rfe\ {i.e., at variance with Pop. A where the majority of sources have measurable \rfe.  A defining feature is the presence of a redward asymmetry, especially prominent in low-ionization lines \citep{marzianietal03b,wangetal17,wolfetal20}. The \rfe\ parameter is a fundamental one in the definition of the MS, and the \feii\ flux is expected to be dependent not only on the gas physical conditions but on the chemical abundances as well. It is therefore important to assess the relevance of metallicity $Z$\ on the location in optical plane of the MS \citep{punslyetal18a,pandaetal18,pandaetal19}.  A second important result coming from the MS occupation is that, at variance with Pop. A,  extreme Pop. B encompasses  the largest prevalence of ``jetted'' sources. It is intriguing that the jetted sources differ from the RQ ones in the distribution of blueshift of high ionization lines, while there is a minor effect of radio-loudness on the low-ionization lines \citep{sulenticetal07,richardsetal11}.   Extreme Pop. A also includes a significant fraction of powerful radio emitters, but their nature is probably very different from the one of the jetted sources in Pop. B, as discussed in \S\ \ref{radio}.   }

\subsection{Chemical composition analysis}
\label{zextremeb}

{ While the MS emission line trends might be mainly governed by a trend in ionization parameter and density, there are other factors that may play a concomitant role, and whose importance is still debated \citep[see e.g.,][]{templeetal21}. One of  them is the metal content of the line emitting gas.} We considered the active galaxy NGC 1275 (a.k.a. 3C 84 or Perseus A, a radio-loud AGN with an optical spectrum of Pop. B), and analyzed the rest-frame optical and UV spectra to measure the  chemical abundance of its BLR gas. {A full account of the analysis is provided by \citet{punslyetal18a}. } The source has a very faint BLR compared to the prominent narrow-line region emission  but, once this difficulty has been overcome via a careful multi-component nonlinear fit, it was possible to gain constraints on the target line flux ratios: \rfe $\lesssim 0.3$,\ {FWHM \hb $\approx$ 5000 \kms\  (implying a spectral type B1 in the MS context)\footnote{The same considerations however apply to extreme Pop. B, considering that those sources show emission line ratios similar to the NGC 1275.}}, low  [C{\sc iv}]$\lambda$1549/He{\sc ii}$\lambda$1640 and C{\sc iv}$\lambda$1549/\hb\ ratios, high He{\sc ii}$\lambda$4686/\hb.  { The fit was carried out using a non-linear, multicomponent, minimum $\chi^2$\	 approach with the same technique described in several previous papers \citep[e.g.,][]{marzianietal10,sulenticetal15}.} { We exploit here an empirical advantage of the MS: for most of the bins it is possible to consider composites or even a single object as representative of the spectral type \citep{marzianietal10}, as  (the only exception being perhaps A1; \citealt{pandaetal19}) the sources scatter around a well defined average.}

{ Recent analyses utilized photoionization computations that cover a six-dimensional parameter space (spectral energy distribution (SED), ionization parameter, density, column density metallicity, micro-turbulence;  \citealt{pandaetal18}).}   Exploration of a parameter subspace in density, ionization parameter U, column density, and  metallicity with CLOUDY 17.01 \citep{ferlandetal17} assuming a carefully defined SED on the basis of multifrequency observations ({ a full description is provided by \citealt{punslyetal18a}}) yielded a set of possible solution  in agreement with the diagnostic ratios, and with  the observed line luminosity: $n_\mathrm{H} \sim 10^{10}$ cm$^{-3}$, $N_\mathrm{H} \sim 10^{23}$ cm$^{-2}$, $0.1 Z_\odot \lesssim Z \lesssim 1 Z_\odot$\ \citep{punslyetal18a}. These values are consistent with the ones assumed by  \citet{pandaetal19} for the Pop. B spectral bins.  

\subsection{Orientation analysis}
\label{orien}

Type-1 quasars showing a Fanaroff-Riley II morphology are associated with very low Eddington ratio and moderate $\theta$. Actually the core-to-lobe ratio and the core-to-optical fluxes have been used as an orientation indicator \citep[][]{willsbrowne86,willsbrotherton95}. In the MS context, Fanaroff-Riley II quasars and also jetted sources \citep[satisfying the rather strict condition on the radio-to-optical flux $\gtrsim 80$][]{padovani17,sulenticetal03}  are constrained within Pop. B, with few core-dominated sources entering the domain of Pop- A (i.e., appearing with FWHM \hb $\lesssim$ 4000 \kms) because of the almost face-on orientation  \citep{sulenticetal03}. { Fig. \ref{fig:inclination} illustrates the appearance of line profiles for two objects of Pop. B with widely different line widths, Arp 102B and the blazar PKS 0438-436. Both objects are fairly well representative of their classes: Arp 102B for double-peaked sources, and PKS 0438-436 for blazars. The middle panel illustrates the displacement expected on the basis of the interpretation scheme outlined in \S \ref{interp}.}

\subsubsection{Moderate-to-high viewing angles}

Sources with widely-separated double peaks attracted a lot of attention in the  years 1990s and 2000s. Very broad Population B sources   that show a double peaked Balmer line profile have been long since considered as candidate accretion disk profiles \citep{chenetal89}. They are rare ($\sim$ 2\% in the SDSS; \citealt{stratevaetal03}) and, after discounting the hypothesis of a binary black hole \citep{halpernfilippenko88} on the basis of long-term monitoring \citep{eracleousetal97},  as well as the possibility of a bipolar outflow \citep{sulenticetal90,zhengetal90} for lack of evidence, the present consensus is that the profiles are genuinely representative of disk emission, possibly without strict axial symmetry \citep{eracleousetal95,bonetal18}. Basically, the radial velocity stability of the double-peaked structure ruled out alternative models. The prototypical source Arp 102B is relatively bright and luminous ($M_\mathrm{R } \approx -25.7$) but nowadays very similar profiles are being discovered in lower luminosity AGN \citep{bianchietal19}, thanks to the careful subtraction of the host galaxy continuum made possible by the use of integral field spectroscopy. Models of accretion disks are consistent with viewing angle around 30 — 50 degrees, and show the red wing with redshift amplitude $\delta z$ consistent with the effect of gravitational and transverse redshift \citep{zhengsulentic90,corbin95,popovicetal95,bonetal15}, i.e.

\begin{equation}
\delta z \approx \frac{3}{2}\frac{r_\mathrm{g}}{\tilde{r}_\mathrm{BLR}}
\end{equation}

where $r_\mathrm{g}$\ is the gravitational radius, and $\tilde{r}_\mathrm{BLR}$\   corresponds to innermost radii of the BLR, if the $\delta z$\ measurement is obtained from the line centroid close to the line base (for example, 0.1 or 0.25 of maximum intensity). 

\subsubsection{Low viewing angles}

The emission   line profiles of radio-loud quasars are often characterized by extreme redward asymmetries in the line profiles \citep{marzianietal96,punsly10}. A recent survey of blazar spectra  \citep{punslyetal20}  show consistency with a large  contribution  from the inner region of an accretion disk to the line profiles, with disk inner radius $\tilde{r}_\mathrm{BLR} \lesssim 100 r_\mathrm{g}$. In this case, the model profiles constrain the disk viewing angle to be $\theta \lesssim 5$ degrees.  The accretion disk emission is  favored by the low level of the ionizing continuum: if normalized to the optical flux, the emission is about two order of magnitude lower in the FUV domain with respect to NGC 5548, the prototypical Population B source.  The flatness of the emitting region added a second element that helped to detect  the effect of orientation, if the blazar profiles are compared to the ones of more inclined sources such as Arp 102B. 

\begin{figure}[htp!]
	\centerline{\includegraphics[ scale=0.5]{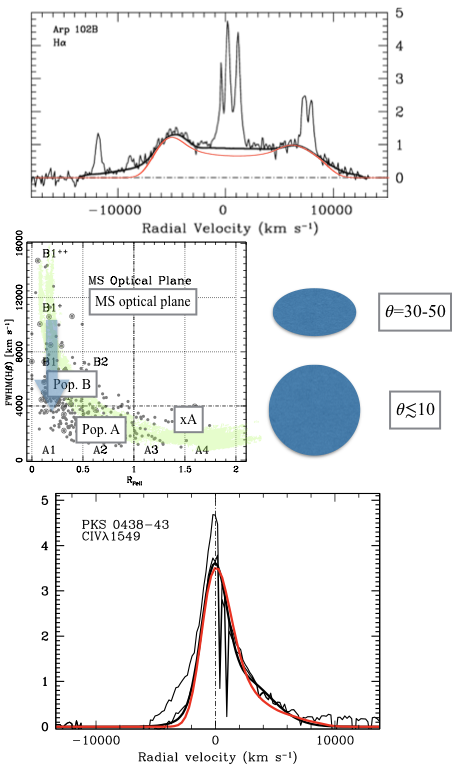}}
	\caption{Effect of the viewing angle on the line profiles of emission lines for \rfe $\sim 0$. The top panel shows the H$\alpha$ profile  of Arp 102B, in the rest frame, after continuum subtraction and  atmospheric absorption correction \citep{sulenticetal90}. The black line shows an empirical fit to the broad emission, while the red one is the relativistic accretion disk model (inner and outer radius $r_\mathrm{in} = 350 r_\mathrm{g}, r_\mathrm{out} = 1000 r_\mathrm{g}, \theta = 32$).  The middle panel is meant to illustrate the effect of a large change in viewing angle $\theta$\ over the location of a source in the optical plane of the MS at \rfe $\sim 0$, in accordance with the prediction of an accretion disk seen at different orientation.  The C{\sc iv } profile of the blazar PKS 0438-436 has been analysed by \citet{punslyetal18}. In this case the disk model has been computed for inner and outer radius $r_\mathrm{in} = 100 r_\mathrm{g}, r_\mathrm{out} \approx 8000 r_\mathrm{g}, \theta = 9$. \label{fig:inclination}}
\end{figure}

The detection of gravitational redshift requires efficient illumination of the inner disk, which may be provided by low luminosity or by particular geometries, for example a warped disk \citep{jiangetal21}.  As a corollary, it is unclear whether redshifted line wings in all of Population B can be explained as due to gravitational and  transverse redshift.  The C{\sc iv}$\lambda$1549 shown in  Fig. \ref{fig:inclination} illustrates the point: while the red wing is well reproduced by a disk model, the core and the blue line sides are affected by additional line emission roughly at rest frame, as well as by an outflowing component yielding an excess emission that voids the information from the centroid close to the line base \citep{popovicetal04,bonetal06,bon08,bonetal09a}.

\section{Extreme Population A}\label{sec3}
\label{extremea}

The extreme Population A  (xA)  is made of  \feii\ emitters with \rfe $\gtrsim 1$\ \citep{marzianisulentic14}.  The CIV$\lambda$1549 is mainly blueshifted (i.e. wind) emission but  Balmer and Paschen lines remain predominantly virialized \citep{martinez-aldamaetal18a}. The selection criterion  \rfe $\gtrsim 1$\ corresponds to  an UV selection criterion based on two diagnostic ratios  implying strong Aluminium and Silicon lines and weak C{\sc iii}]$\lambda$1909.  At the origin of the interest in xA sources (some of which might be highly super-Eddington accretors, \citealt{wangetal14b}) is the fact that xA population quasars radiate close to an extreme  Eddington ratio \citep{marzianisulentic14}, and  this property may be exploited to define  standard “Eddington candles” for cosmology \citep[e.g.,][ and references therein]{marzianietal21}. 

\subsection{Chemical composition analysis}

{ Emission line ratios in a photoionization context are dependent on the SED of the ionising radiation that is known in turn to depend on Eddington ratio \citep{laoretal97b,ferlandetal20}, and hence on the location along the MS. } Arrays of Cloudy 17.02 \citep{ferlandetal17}  { photoionization} simulations were computed covering the $U$ — density  parameter plane with a step of 0.25 dex, for 12 values of metallicity covering the range 0.01 $Z_\odot \le Z \le 1000   Z_\odot$, { with a SED appropriate for sources radiating at high Eddington ratio, including a prominent big blue bump}. A detailed account of the simulation settings and a systematic presentation { the physical basis of the method and } of the results is provided by \citet{marzianietal20}\ and \citet{sniegowskaetal21}. The similation results   show that diagnostic line ratios  CIV$\lambda$1549/He{\sc ii}$\lambda$1640,      Al{\sc iii} $\lambda$1860/He{\sc ii}$\lambda$1640,  (Si{\sc iv}+O{\sc iv}])$\lambda$1400/He{\sc ii}$\lambda$1640   are monotonically increasing with $Z$ over a wide range of ionization parameter values, for a fixed SED  \citep{sniegowskaetal21}. The He{\sc ii}$\lambda$1640 emission line   is expedient because of unchanging He abundance and of simple He{\sc ii}$\lambda$1640 radiation transfer (collisional excitation is negligible, as the lower transition level is at a high energy above ground $\approx 40$eV; \citealt{marzianietal20}).  

In terms of metallicity,  xA sources show very homogeneous properties  and extreme  values of metallicity ($Z\gtrsim 10 Z_\odot$; $Z \sim 20 Z_\odot$\ seems a typical value).  For the low-ionizaton BLR, ionization parameter, density, and $Z $\ are constrained within a relatively narrow range for most xA sources \citep{sniegowskaetal21}.  Systematic differences in the $Z$\ derived from the different ratios may imply over-abundances of Al and Si ($Z$(Al{\sc iii}/He{\sc ii})$ \sim 20 - 50 Z_\odot$) with respect to C, since  $Z$(CIV/He{\sc ii}) $\sim 10 Z_\odot$ (Garnica et al. 2021, in preparation). The  difference persists over a rather wide range in ionization parameter and density, and is most likely due to scaling by a fixed factor under the assumption that {\em all}  relative elemental abundances of the BLR are solar. This approach does not take into account that  solar relative abundances are most likely inappropriate in the nuclear and circumnucler regions of AGN. 

\subsection{Radio properties}
\label{radio}

An unexpected finding was   a high prevalence of radio-intermediate (with radio-to-optical specific flux in the range 10 — 80)  or even radio-loud quasars in xA that can reach  very high radio power $P\nu \lesssim 10^{25}$ W Hz$^{-1}$, comparable to the  low end of the  distribution of jetted sources \citep{zamfiretal08}. Moving along the sequence from spectral type from B1$^{++}$ to A4, the fraction of  radio-loud and radio-intermediate   peaks at the two extremes. 

\begin{figure}[t]
	\centerline{\includegraphics[ scale=0.45]{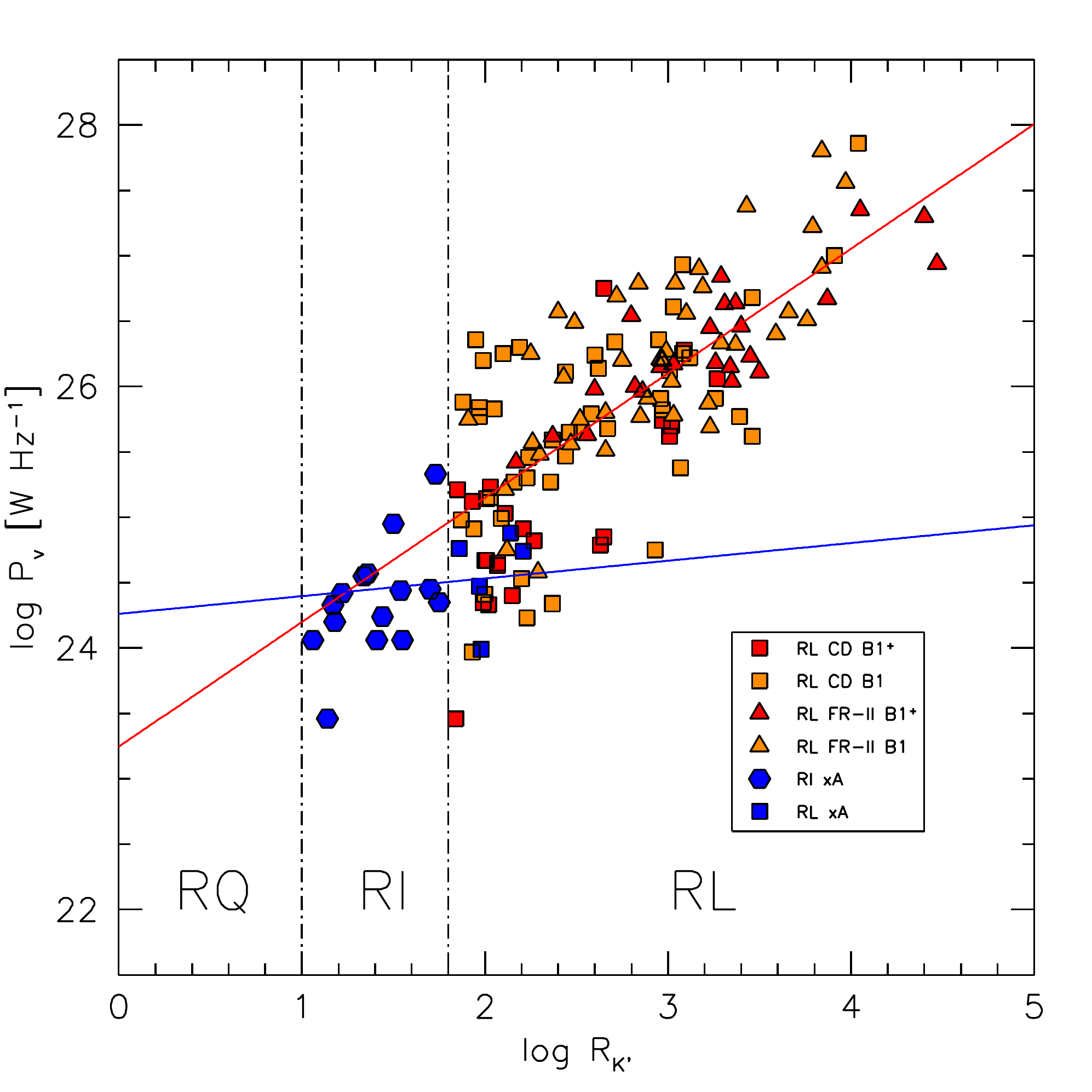}}
	\caption{Relation between radio power and radio-to-optical flux ratio for the xA sources  and for jetted sources of Pop. B  spectral types. Squares and triangles: compact and FR-II radio-loud; hexagons: radio-intermediate. Red: spectral type B1$^+$, orange: B1; navy blue: xA represented by 	spectral type A3 and A4, { squares for RL, and exagons for radio intermediate (RI).   The dot-dashed lines separate the domain of RQ, RI, and RL Pop. B. The blue and red lines are an unweighted lsq fit for the xA sources} and for the RL Pop. B objects, respectively. \label{fig:rlri}}
\end{figure}

A possible interpretation  is the existence of a population of jetted sources of moderate luminosity, associated with small black hole masses  \citep[RL NLSy1s, ][]{komossaetal06}. However, the xA radio-intermediate { might} not follow the same relation between radio power and radio-to-optical flux ratio (Fig. \ref{fig:rlri}, and \citealt{delolmoetal21}). WISE colors indicate that star formation in the host galaxies can give a significant contribution to the radio power  \citep{caccianigaetal15}.   \citet{gancietal19} showed that the xA sources with significantly radio power obey the correlation between FIR luminosity and radio power expected for star-forming galaxies and radio-quiet quasars  \citep[c.f.][]{bonzinietal15}. A radio power $P\nu \sim 10^{25}$ W Hz$^{-1}$\ translates into an enormous star-formation rate SFR $\sim 10^4$ M$_\odot$ yr$^{-1}$, implying that most of the host (and not only the circumnuclear regions) might be experiencing a burst of star formation. This might be unlikely at low redshift, but not at the “cosmic noon.”

\subsection{Orientation effects}

{ The orientation effects on xA sources specifically have been analyzed in several recent papers \citep{negreteetal18,dultzinetal20,marzianietal20a}. xA samples are  biased: since they have low line equivalent width \citep{martinez-aldamaetal18a}, they are preferentially selected with narrower lines, and narrower lines might imply low inclination. This said,  xA sources are accreting at very high rates,  implying that their luminosity per unit mass converges toward a limiting value \citep{wangetal13}. This property can be, in principle, used to define  “Eddington standard candles,” in which the Eddington ratio, and not the luminosity, has a small scatter around a well-defined value \citep{marzianisulentic14}.   Since the virial luminosity is derived from the assumption $L/M_\mathrm{BH} \sim const$, lower $L$ is implied by narrower line width, and the FWHM decreases with decreasing amplitude of the viewing angle,  a relation can be established between viewing angle $\theta$\ and the difference between the virial  luminosity and luminosity derived from concordance cosmology. The resulting distribution covers the range from $0 \lesssim \theta \lesssim 50$, with a maximum  at $\theta \approx 20$  \citep{negreteetal18,marzianietal20a},  with a relatively small dispersion. At least the xA sources that were analyzed  in the previous works are seen predominantly almost face-on. }




\section{Conclusion}
\label{concl}

Recent developments strengthen the interpretation of the MS of quasars  based on Eddington ratio and orientation. At the Population B extremes, the broad line emitting regions appear as  “disk dominated” and the flatness of the emitting region helped detect orientation effects. The MS is not only about  spectral parameters; instead, it reflects different evolutionary and environmental situations. In Population B, an estimate of metallicity $Z$\ for a fairly typical source suggest solar or slightly sub-solar metallicity.  At the other extreme of the main sequence,  extreme of Population A appear to be in highly star forming hosts,  very metal rich, possibly with enrichment associated with a circumnuclear Starburst or nuclear accretion modified stars \citep{donofriomarziani18}. Preliminary attempts to build a Hubble diagram with xA sources are still affected by large dispersion in the distance modulus estimates  \citep{dultzinetal20,czernyetal21}. Large samples (ideally recognized  via machine learning, \citealt{jankovetal21,peruzzietal21} could help) reduce the statistical dispersion. However, the  additional complexity related to SED and  circumnuclear environment needs to be thoroughly analyzed to safely exploit  xA  sources   as Eddington standard candles.

\subsection*{Author contributions}

PM wrote the paper.  The other authors contributed in various form  to several publications summarized in the text. { We thank the reviewer whose suggestion greatly contributed to improve the clarity and completeness of the paper. }

\subsection*{Financial disclosure}

NB and EB acknowledge the support of Serbian Ministry of Education, Science and Technological Development, through the contract number 451-03-68/2020-14/200002.


\begin{thebibliography}{}

\bibitem [\protect \citeauthoryear {%
{Antonucci}%
}{%
{Antonucci}%
}{%
{\protect \APACyear {1993}}%
}]{%
antonucci93}
\APACinsertmetastar {%
antonucci93}%
\begin{APACrefauthors}%
{Antonucci}, R.%
\end{APACrefauthors}%
\unskip\
\newblock
\APACrefYearMonthDay{1993}{}{},
\newblock
\unskip
\newblock
\APACjournalVolNumPages{\araa}{31}{}{473-521}.
\newblock
\begin{APACrefDOI} \doi{10.1146/annurev.aa.31.090193.002353} \end{APACrefDOI}
\PrintBackRefs{\CurrentBib}

\bibitem [\protect \citeauthoryear {%
{Bentz}%
\ \protect \BOthers {.}}{%
{Bentz}%
\ \protect \BOthers {.}}{%
{\protect \APACyear {2013}}%
}]{%
bentzetal13}
\APACinsertmetastar {%
bentzetal13}%
\begin{APACrefauthors}%
{Bentz}, M\BPBI C.%
, {Denney}, K\BPBI D.%
, {Grier}, C\BPBI J.%
\ et al.\end{APACrefauthors}%
\unskip\
\newblock
\APACrefYearMonthDay{2013}{{\APACmonth{04}}}{},
\newblock
\unskip
\newblock
\APACjournalVolNumPages{\apj}{767}{}{149}.
\newblock
\begin{APACrefDOI} \doi{10.1088/0004-637X/767/2/149} \end{APACrefDOI}
\PrintBackRefs{\CurrentBib}

\bibitem [\protect \citeauthoryear {%
{Bianchi}%
\ \protect \BOthers {.}}{%
{Bianchi}%
\ \protect \BOthers {.}}{%
{\protect \APACyear {2019}}%
}]{%
bianchietal19}
\APACinsertmetastar {%
bianchietal19}%
\begin{APACrefauthors}%
{Bianchi}, S.%
, {Antonucci}, R.%
, {Capetti}, A.%
\ et al.\end{APACrefauthors}%
\unskip\
\newblock
\APACrefYearMonthDay{2019}{{\APACmonth{09}}}{},
\newblock
\unskip
\newblock
\APACjournalVolNumPages{\mnras}{488}{1}{L1-L5}.
\newblock
\begin{APACrefDOI} \doi{10.1093/mnrasl/slz080} \end{APACrefDOI}
\PrintBackRefs{\CurrentBib}

\bibitem [\protect \citeauthoryear {%
E.~{Bon}%
}{%
E.~{Bon}%
}{%
{\protect \APACyear {2008}}%
}]{%
bon08}
\APACinsertmetastar {%
bon08}%
\begin{APACrefauthors}%
{Bon}, E.%
\end{APACrefauthors}%
\unskip\
\newblock
\APACrefYearMonthDay{2008}{{\APACmonth{12}}}{},
\newblock
\unskip
\newblock
\APACjournalVolNumPages{Serbian Astronomical Journal}{177}{}{9-13}.
\newblock
\begin{APACrefDOI} \doi{10.2298/SAJ0877009B} \end{APACrefDOI}
\PrintBackRefs{\CurrentBib}

\bibitem [\protect \citeauthoryear {%
E.~{Bon}%
, {Jovanovi{\'c}}%
, {Marziani}%
, {Bon}%
\BCBL {}\ \BBA {} {Ota{\v{s}}evi{\'c}}%
}{%
E.~{Bon}%
\ \protect \BOthers {.}}{%
{\protect \APACyear {2018}}%
}]{%
bonetal18}
\APACinsertmetastar {%
bonetal18}%
\begin{APACrefauthors}%
{Bon}, E.%
, {Jovanovi{\'c}}, P.%
, {Marziani}, P.%
, {Bon}, N.%
\BCBL {}\ \BBA {} {Ota{\v{s}}evi{\'c}}, A.%
\end{APACrefauthors}%
\unskip\
\newblock
\APACrefYearMonthDay{2018}{{\APACmonth{06}}}{},
\newblock
\unskip
\newblock
\APACjournalVolNumPages{Frontiers in Astronomy and Space Sciences}{5}{}{19}.
\newblock
\begin{APACrefDOI} \doi{10.3389/fspas.2018.00019} \end{APACrefDOI}
\PrintBackRefs{\CurrentBib}

\bibitem [\protect \citeauthoryear {%
E.~{Bon}%
, {Popovi{\'c}}%
, {Gavrilovi{\'c}}%
, {Mura}%
\BCBL {}\ \BBA {} {Mediavilla}%
}{%
E.~{Bon}%
\ \protect \BOthers {.}}{%
{\protect \APACyear {2009}}%
}]{%
bonetal09a}
\APACinsertmetastar {%
bonetal09a}%
\begin{APACrefauthors}%
{Bon}, E.%
, {Popovi{\'c}}, L\BPBI {\v C}.%
, {Gavrilovi{\'c}}, N.%
, {Mura}, G\BPBI L.%
\BCBL {}\ \BBA {} {Mediavilla}, E.%
\end{APACrefauthors}%
\unskip\
\newblock
\APACrefYearMonthDay{2009}{{\APACmonth{12}}}{},
\newblock
\unskip
\newblock
\APACjournalVolNumPages{\mnras}{400}{}{924-936}.
\newblock
\begin{APACrefDOI} \doi{10.1111/j.1365-2966.2009.15511.x} \end{APACrefDOI}
\PrintBackRefs{\CurrentBib}

\bibitem [\protect \citeauthoryear {%
E.~{Bon}%
, {Popovi{\'c}}%
, {Ili{\'c}}%
\BCBL {}\ \BBA {} {Mediavilla}%
}{%
E.~{Bon}%
\ \protect \BOthers {.}}{%
{\protect \APACyear {2006}}%
}]{%
bonetal06}
\APACinsertmetastar {%
bonetal06}%
\begin{APACrefauthors}%
{Bon}, E.%
, {Popovi{\'c}}, L\BPBI {\v{C}}.%
, {Ili{\'c}}, D.%
\BCBL {}\ \BBA {} {Mediavilla}, E.%
\end{APACrefauthors}%
\unskip\
\newblock
\APACrefYearMonthDay{2006}{{\APACmonth{11}}}{},
\newblock
\unskip
\newblock
\APACjournalVolNumPages{\nar}{50}{9-10}{716-719}.
\newblock
\begin{APACrefDOI} \doi{10.1016/j.newar.2006.06.015} \end{APACrefDOI}
\PrintBackRefs{\CurrentBib}

\bibitem [\protect \citeauthoryear {%
N.~{Bon}%
, {Bon}%
\BCBL {}\ \BBA {} {Marziani}%
}{%
N.~{Bon}%
\ \protect \BOthers {.}}{%
{\protect \APACyear {2018}}%
}]{%
bonetal18a}
\APACinsertmetastar {%
bonetal18a}%
\begin{APACrefauthors}%
{Bon}, N.%
, {Bon}, E.%
\BCBL {}\ \BBA {} {Marziani}, P.%
\end{APACrefauthors}%
\unskip\
\newblock
\APACrefYearMonthDay{2018}{{\APACmonth{01}}}{},
\newblock
\unskip
\newblock
\APACjournalVolNumPages{Frontiers in Astronomy and Space Sciences}{5}{}{3}.
\newblock
\begin{APACrefDOI} \doi{10.3389/fspas.2018.00003} \end{APACrefDOI}
\PrintBackRefs{\CurrentBib}

\bibitem [\protect \citeauthoryear {%
N.~{Bon}%
, {Bon}%
, {Marziani}%
\BCBL {}\ \BBA {} {Jovanovi{\'c}}%
}{%
N.~{Bon}%
\ \protect \BOthers {.}}{%
{\protect \APACyear {2015}}%
}]{%
bonetal15}
\APACinsertmetastar {%
bonetal15}%
\begin{APACrefauthors}%
{Bon}, N.%
, {Bon}, E.%
, {Marziani}, P.%
\BCBL {}\ \BBA {} {Jovanovi{\'c}}, P.%
\end{APACrefauthors}%
\unskip\
\newblock
\APACrefYearMonthDay{2015}{{\APACmonth{12}}}{},
\newblock
\unskip
\newblock
\APACjournalVolNumPages{\apss}{360}{}{7}.
\newblock
\begin{APACrefDOI} \doi{10.1007/s10509-015-2555-5} \end{APACrefDOI}
\PrintBackRefs{\CurrentBib}

\bibitem [\protect \citeauthoryear {%
N.~{Bon}%
\ \protect \BOthers {.}}{%
N.~{Bon}%
\ \protect \BOthers {.}}{%
{\protect \APACyear {2020}}%
}]{%
bonetal20}
\APACinsertmetastar {%
bonetal20}%
\begin{APACrefauthors}%
{Bon}, N.%
, {Marziani}, P.%
, {Bon}, E.%
\ et al.\end{APACrefauthors}%
\unskip\
\newblock
\APACrefYearMonthDay{2020}{{\APACmonth{03}}}{},
\newblock
\unskip
\newblock
\APACjournalVolNumPages{\aap}{635}{}{A151}.
\newblock
\begin{APACrefDOI} \doi{10.1051/0004-6361/201936773} \end{APACrefDOI}
\PrintBackRefs{\CurrentBib}

\bibitem [\protect \citeauthoryear {%
{Bonzini}%
\ \protect \BOthers {.}}{%
{Bonzini}%
\ \protect \BOthers {.}}{%
{\protect \APACyear {2015}}%
}]{%
bonzinietal15}
\APACinsertmetastar {%
bonzinietal15}%
\begin{APACrefauthors}%
{Bonzini}, M.%
, {Mainieri}, V.%
, {Padovani}, P.%
\ et al.\end{APACrefauthors}%
\unskip\
\newblock
\APACrefYearMonthDay{2015}{{\APACmonth{10}}}{},
\newblock
\unskip
\newblock
\APACjournalVolNumPages{\mnras}{453}{}{1079-1094}.
\newblock
\begin{APACrefDOI} \doi{10.1093/mnras/stv1675} \end{APACrefDOI}
\PrintBackRefs{\CurrentBib}

\bibitem [\protect \citeauthoryear {%
{Boroson}%
\ \BBA {} {Green}%
}{%
{Boroson}%
\ \BBA {} {Green}%
}{%
{\protect \APACyear {1992}}%
}]{%
borosongreen92}
\APACinsertmetastar {%
borosongreen92}%
\begin{APACrefauthors}%
{Boroson}, T\BPBI A.%
\BCBT {}\ \BBA {} {Green}, R\BPBI F.%
\end{APACrefauthors}%
\unskip\
\newblock
\APACrefYearMonthDay{1992}{{\APACmonth{05}}}{},
\newblock
\unskip
\newblock
\APACjournalVolNumPages{\apjs}{80}{}{109}.
\newblock
\begin{APACrefDOI} \doi{10.1086/191661} \end{APACrefDOI}
\PrintBackRefs{\CurrentBib}

\bibitem [\protect \citeauthoryear {%
{Caccianiga}%
\ \protect \BOthers {.}}{%
{Caccianiga}%
\ \protect \BOthers {.}}{%
{\protect \APACyear {2015}}%
}]{%
caccianigaetal15}
\APACinsertmetastar {%
caccianigaetal15}%
\begin{APACrefauthors}%
{Caccianiga}, A.%
, {Ant{\'o}n}, S.%
, {Ballo}, L.%
\ et al.\end{APACrefauthors}%
\unskip\
\newblock
\APACrefYearMonthDay{2015}{{\APACmonth{08}}}{},
\newblock
\unskip
\newblock
\APACjournalVolNumPages{\mnras}{451}{2}{1795-1805}.
\newblock
\begin{APACrefDOI} \doi{10.1093/mnras/stv939} \end{APACrefDOI}
\PrintBackRefs{\CurrentBib}

\bibitem [\protect \citeauthoryear {%
{Chen}%
, {Halpern}%
\BCBL {}\ \BBA {} {Filippenko}%
}{%
{Chen}%
\ \protect \BOthers {.}}{%
{\protect \APACyear {1989}}%
}]{%
chenetal89}
\APACinsertmetastar {%
chenetal89}%
\begin{APACrefauthors}%
{Chen}, K.%
, {Halpern}, J\BPBI P.%
\BCBL {}\ \BBA {} {Filippenko}, A\BPBI V.%
\end{APACrefauthors}%
\unskip\
\newblock
\APACrefYearMonthDay{1989}{{\APACmonth{04}}}{},
\newblock
\unskip
\newblock
\APACjournalVolNumPages{\apj}{339}{}{742-751}.
\newblock
\begin{APACrefDOI} \doi{10.1086/167332} \end{APACrefDOI}
\PrintBackRefs{\CurrentBib}

\bibitem [\protect \citeauthoryear {%
{Collin-Souffrin}%
, {Dyson}%
, {McDowell}%
\BCBL {}\ \BBA {} {Perry}%
}{%
{Collin-Souffrin}%
\ \protect \BOthers {.}}{%
{\protect \APACyear {1988}}%
}]{%
collinsouffrinetal88}
\APACinsertmetastar {%
collinsouffrinetal88}%
\begin{APACrefauthors}%
{Collin-Souffrin}, S.%
, {Dyson}, J\BPBI E.%
, {McDowell}, J\BPBI C.%
\BCBL {}\ \BBA {} {Perry}, J\BPBI J.%
\end{APACrefauthors}%
\unskip\
\newblock
\APACrefYearMonthDay{1988}{{\APACmonth{06}}}{},
\newblock
\unskip
\newblock
\APACjournalVolNumPages{MNRAS}{232}{}{539-550}.
\PrintBackRefs{\CurrentBib}

\bibitem [\protect \citeauthoryear {%
{Corbin}%
}{%
{Corbin}%
}{%
{\protect \APACyear {1995}}%
}]{%
corbin95}
\APACinsertmetastar {%
corbin95}%
\begin{APACrefauthors}%
{Corbin}, M\BPBI R.%
\end{APACrefauthors}%
\unskip\
\newblock
\APACrefYearMonthDay{1995}{{\APACmonth{07}}}{},
\newblock
\unskip
\newblock
\APACjournalVolNumPages{\apj}{447}{}{496-+}.
\newblock
\begin{APACrefDOI} \doi{10.1086/175894} \end{APACrefDOI}
\PrintBackRefs{\CurrentBib}

\bibitem [\protect \citeauthoryear {%
{Czerny}%
\ \protect \BOthers {.}}{%
{Czerny}%
\ \protect \BOthers {.}}{%
{\protect \APACyear {2021}}%
}]{%
czernyetal21}
\APACinsertmetastar {%
czernyetal21}%
\begin{APACrefauthors}%
{Czerny}, B.%
, {Mart{\'\i}nez-Aldama}, M\BPBI L.%
, {Wojtkowska}, G.%
\ et al.\end{APACrefauthors}%
\unskip\
\newblock
\APACrefYearMonthDay{2021}{{\APACmonth{04}}}{},
\newblock
\unskip
\newblock
\APACjournalVolNumPages{Acta Physica Polonica A}{139}{4}{389-393}.
\newblock
\begin{APACrefDOI} \doi{10.12693/APhysPolA.139.389} \end{APACrefDOI}
\PrintBackRefs{\CurrentBib}

\bibitem [\protect \citeauthoryear {%
{del Olmo}%
\ \protect \BOthers {.}}{%
{del Olmo}%
\ \protect \BOthers {.}}{%
{\protect \APACyear {2021}}%
}]{%
delolmoetal21}
\APACinsertmetastar {%
delolmoetal21}%
\begin{APACrefauthors}%
{del Olmo}, A.%
, {Marziani}, P.%
, {Ganci}, V.%
, {D'Onofrio}, M.%
, {Bon}, E.%
, {Bon}, N.%
\BCBL {}\ \BBA {} {Negrete}, A\BPBI C.%
\end{APACrefauthors}%
\unskip\
\newblock
\APACrefYearMonthDay{2021}{{\APACmonth{01}}}{},
\newblock
\unskip
\newblock
\APACjournalVolNumPages{IAU Symposium}{356}{}{310-313}.
\newblock
\begin{APACrefDOI} \doi{10.1017/S1743921320003191} \end{APACrefDOI}
\PrintBackRefs{\CurrentBib}

\bibitem [\protect \citeauthoryear {%
{D'Onofrio}%
\ \BBA {} {Marziani}%
}{%
{D'Onofrio}%
\ \BBA {} {Marziani}%
}{%
{\protect \APACyear {2018}}%
}]{%
donofriomarziani18}
\APACinsertmetastar {%
donofriomarziani18}%
\begin{APACrefauthors}%
{D'Onofrio}, M.%
\BCBT {}\ \BBA {} {Marziani}, P.%
\end{APACrefauthors}%
\unskip\
\newblock
\APACrefYearMonthDay{2018}{{\APACmonth{09}}}{},
\newblock
\unskip
\newblock
\APACjournalVolNumPages{Frontiers in Astronomy and Space Sciences}{5}{}{31}.
\newblock
\begin{APACrefDOI} \doi{10.3389/fspas.2018.00031} \end{APACrefDOI}
\PrintBackRefs{\CurrentBib}

\bibitem [\protect \citeauthoryear {%
{D'Onofrio}%
, {Marziani}%
\BCBL {}\ \BBA {} {Chiosi}%
}{%
{D'Onofrio}%
\ \protect \BOthers {.}}{%
{\protect \APACyear {2021}}%
}]{%
donofrioetal21}
\APACinsertmetastar {%
donofrioetal21}%
\begin{APACrefauthors}%
{D'Onofrio}, M.%
, {Marziani}, P.%
\BCBL {}\ \BBA {} {Chiosi}, C.%
\end{APACrefauthors}%
\unskip\
\newblock
\APACrefYearMonthDay{2021}{{\APACmonth{09}}}{},
\newblock
\unskip
\newblock
\APACjournalVolNumPages{arXiv e-prints}{}{}{arXiv:2109.06301}.
\PrintBackRefs{\CurrentBib}

\bibitem [\protect \citeauthoryear {%
{Du}%
\ \protect \BOthers {.}}{%
{Du}%
\ \protect \BOthers {.}}{%
{\protect \APACyear {2016}}%
}]{%
duetal16a}
\APACinsertmetastar {%
duetal16a}%
\begin{APACrefauthors}%
{Du}, P.%
, {Wang}, J\BHBI M.%
, {Hu}, C.%
, {Ho}, L\BPBI C.%
, {Li}, Y\BHBI R.%
\BCBL {}\ \BBA {} {Bai}, J\BHBI M.%
\end{APACrefauthors}%
\unskip\
\newblock
\APACrefYearMonthDay{2016}{{\APACmonth{02}}}{},
\newblock
\unskip
\newblock
\APACjournalVolNumPages{\apjl}{818}{}{L14}.
\newblock
\begin{APACrefDOI} \doi{10.3847/2041-8205/818/1/L14} \end{APACrefDOI}
\PrintBackRefs{\CurrentBib}

\bibitem [\protect \citeauthoryear {%
{Dultzin}%
\ \protect \BOthers {.}}{%
{Dultzin}%
\ \protect \BOthers {.}}{%
{\protect \APACyear {2020}}%
}]{%
dultzinetal20}
\APACinsertmetastar {%
dultzinetal20}%
\begin{APACrefauthors}%
{Dultzin}, D.%
, {Marziani}, P.%
, {de Diego}, J\BPBI A.%
\ et al.\end{APACrefauthors}%
\unskip\
\newblock
\APACrefYearMonthDay{2020}{{\APACmonth{01}}}{},
\newblock
\unskip
\newblock
\APACjournalVolNumPages{Frontiers in Astronomy and Space Sciences}{6}{}{80}.
\newblock
\begin{APACrefDOI} \doi{10.3389/fspas.2019.00080} \end{APACrefDOI}
\PrintBackRefs{\CurrentBib}

\bibitem [\protect \citeauthoryear {%
{Elvis}%
}{%
{Elvis}%
}{%
{\protect \APACyear {2000}}%
}]{%
elvis00}
\APACinsertmetastar {%
elvis00}%
\begin{APACrefauthors}%
{Elvis}, M.%
\end{APACrefauthors}%
\unskip\
\newblock
\APACrefYearMonthDay{2000}{{\APACmonth{12}}}{},
\newblock
\unskip
\newblock
\APACjournalVolNumPages{\apj}{545}{}{63-76}.
\newblock
\begin{APACrefDOI} \doi{10.1086/317778} \end{APACrefDOI}
\PrintBackRefs{\CurrentBib}

\bibitem [\protect \citeauthoryear {%
{Eracleous}%
, {Halpern}%
, {Gilbert}%
, {Newman}%
\BCBL {}\ \BBA {} {Filippenko}%
}{%
{Eracleous}%
\ \protect \BOthers {.}}{%
{\protect \APACyear {1997}}%
}]{%
eracleousetal97}
\APACinsertmetastar {%
eracleousetal97}%
\begin{APACrefauthors}%
{Eracleous}, M.%
, {Halpern}, J\BPBI P.%
, {Gilbert}, A\BPBI M.%
, {Newman}, J\BPBI A.%
\BCBL {}\ \BBA {} {Filippenko}, A\BPBI V.%
\end{APACrefauthors}%
\unskip\
\newblock
\APACrefYearMonthDay{1997}{{\APACmonth{11}}}{},
\newblock
\unskip
\newblock
\APACjournalVolNumPages{\apj}{490}{}{216-+}.
\newblock
\begin{APACrefDOI} \doi{10.1086/304859} \end{APACrefDOI}
\PrintBackRefs{\CurrentBib}

\bibitem [\protect \citeauthoryear {%
{Eracleous}%
, {Livio}%
, {Halpern}%
\BCBL {}\ \BBA {} {Storchi-Bergmann}%
}{%
{Eracleous}%
\ \protect \BOthers {.}}{%
{\protect \APACyear {1995}}%
}]{%
eracleousetal95}
\APACinsertmetastar {%
eracleousetal95}%
\begin{APACrefauthors}%
{Eracleous}, M.%
, {Livio}, M.%
, {Halpern}, J\BPBI P.%
\BCBL {}\ \BBA {} {Storchi-Bergmann}, T.%
\end{APACrefauthors}%
\unskip\
\newblock
\APACrefYearMonthDay{1995}{{\APACmonth{01}}}{},
\newblock
\unskip
\newblock
\APACjournalVolNumPages{\apj}{438}{}{610-622}.
\newblock
\begin{APACrefDOI} \doi{10.1086/175104} \end{APACrefDOI}
\PrintBackRefs{\CurrentBib}

\bibitem [\protect \citeauthoryear {%
{Ferland}%
\ \protect \BOthers {.}}{%
{Ferland}%
\ \protect \BOthers {.}}{%
{\protect \APACyear {2017}}%
}]{%
ferlandetal17}
\APACinsertmetastar {%
ferlandetal17}%
\begin{APACrefauthors}%
{Ferland}, G\BPBI J.%
, {Chatzikos}, M.%
, {Guzm{\'a}n}, F.%
\ et al.\end{APACrefauthors}%
\unskip\
\newblock
\APACrefYearMonthDay{2017}{{\APACmonth{10}}}{},
\newblock
\unskip
\newblock
\APACjournalVolNumPages{\rmxaa}{53}{}{385-438}.
\PrintBackRefs{\CurrentBib}

\bibitem [\protect \citeauthoryear {%
{Ferland}%
, {Done}%
, {Jin}%
, {Landt}%
\BCBL {}\ \BBA {} {Ward}%
}{%
{Ferland}%
\ \protect \BOthers {.}}{%
{\protect \APACyear {2020}}%
}]{%
ferlandetal20}
\APACinsertmetastar {%
ferlandetal20}%
\begin{APACrefauthors}%
{Ferland}, G\BPBI J.%
, {Done}, C.%
, {Jin}, C.%
, {Landt}, H.%
\BCBL {}\ \BBA {} {Ward}, M\BPBI J.%
\end{APACrefauthors}%
\unskip\
\newblock
\APACrefYearMonthDay{2020}{{\APACmonth{05}}}{},
\newblock
\unskip
\newblock
\APACjournalVolNumPages{\mnras}{494}{4}{5917-5922}.
\newblock
\begin{APACrefDOI} \doi{10.1093/mnras/staa1207} \end{APACrefDOI}
\PrintBackRefs{\CurrentBib}

\bibitem [\protect \citeauthoryear {%
Fraix-Burnet%
, Marziani%
, D'Onofrio%
\BCBL {}\ \BBA {} Dultzin%
}{%
Fraix-Burnet%
\ \protect \BOthers {.}}{%
{\protect \APACyear {2017}}%
}]{%
fraix-burnetetal17}
\APACinsertmetastar {%
fraix-burnetetal17}%
\begin{APACrefauthors}%
Fraix-Burnet, D.%
, Marziani, P.%
, D'Onofrio, M.%
\BCBL {}\ \BBA {} Dultzin, D.%
\end{APACrefauthors}%
\unskip\
\newblock
\APACrefYearMonthDay{2017}{}{},
\newblock
\unskip
\newblock
\APACjournalVolNumPages{Frontiers in Astronomy and Space Sciences}{4}{}{1}.
\newblock
\begin{APACrefURL}
  \url{http://journal.frontiersin.org/article/10.3389/fspas.2017.00001}
  \end{APACrefURL}
\newblock
\begin{APACrefDOI} \doi{10.3389/fspas.2017.00001} \end{APACrefDOI}
\PrintBackRefs{\CurrentBib}

\bibitem [\protect \citeauthoryear {%
{Ganci}%
\ \protect \BOthers {.}}{%
{Ganci}%
\ \protect \BOthers {.}}{%
{\protect \APACyear {2019}}%
}]{%
gancietal19}
\APACinsertmetastar {%
gancietal19}%
\begin{APACrefauthors}%
{Ganci}, V.%
, {Marziani}, P.%
, {D'Onofrio}, M.%
, {del Olmo}, A.%
, {Bon}, E.%
, {Bon}, N.%
\BCBL {}\ \BBA {} {Negrete}, C\BPBI A.%
\end{APACrefauthors}%
\unskip\
\newblock
\APACrefYearMonthDay{2019}{{\APACmonth{10}}}{},
\newblock
\unskip
\newblock
\APACjournalVolNumPages{\aap}{630}{}{A110}.
\newblock
\begin{APACrefDOI} \doi{10.1051/0004-6361/201936270} \end{APACrefDOI}
\PrintBackRefs{\CurrentBib}

\bibitem [\protect \citeauthoryear {%
{Gaskell}%
}{%
{Gaskell}%
}{%
{\protect \APACyear {1985}}%
}]{%
gaskell85}
\APACinsertmetastar {%
gaskell85}%
\begin{APACrefauthors}%
{Gaskell}, C\BPBI M.%
\end{APACrefauthors}%
\unskip\
\newblock
\APACrefYearMonthDay{1985}{{\APACmonth{04}}}{},
\newblock
\unskip
\newblock
\APACjournalVolNumPages{\apj}{291}{}{112-116}.
\newblock
\begin{APACrefDOI} \doi{10.1086/163045} \end{APACrefDOI}
\PrintBackRefs{\CurrentBib}

\bibitem [\protect \citeauthoryear {%
{Halpern}%
\ \BBA {} {Filippenko}%
}{%
{Halpern}%
\ \BBA {} {Filippenko}%
}{%
{\protect \APACyear {1988}}%
}]{%
halpernfilippenko88}
\APACinsertmetastar {%
halpernfilippenko88}%
\begin{APACrefauthors}%
{Halpern}, J\BPBI P.%
\BCBT {}\ \BBA {} {Filippenko}, A\BPBI V.%
\end{APACrefauthors}%
\unskip\
\newblock
\APACrefYearMonthDay{1988}{{\APACmonth{01}}}{},
\newblock
\unskip
\newblock
\APACjournalVolNumPages{Nature}{331}{}{46-48}.
\newblock
\begin{APACrefDOI} \doi{10.1038/331046a0} \end{APACrefDOI}
\PrintBackRefs{\CurrentBib}

\bibitem [\protect \citeauthoryear {%
{Jankov}%
, {Ili{\'c}}%
\BCBL {}\ \BBA {} {Kova{\v{c}}evi{\'c}}%
}{%
{Jankov}%
\ \protect \BOthers {.}}{%
{\protect \APACyear {2021}}%
}]{%
jankovetal21}
\APACinsertmetastar {%
jankovetal21}%
\begin{APACrefauthors}%
{Jankov}, I.%
, {Ili{\'c}}, D.%
\BCBL {}\ \BBA {} {Kova{\v{c}}evi{\'c}}, A.%
\end{APACrefauthors}%
\unskip\
\newblock
\APACrefYearMonthDay{2021}{{\APACmonth{06}}}{},
\newblock
\unskip
\newblock
\APACjournalVolNumPages{Publications de l'Observatoire Astronomique de
  Beograd}{100}{}{241-246}.
\PrintBackRefs{\CurrentBib}

\bibitem [\protect \citeauthoryear {%
{Jiang}%
\ \protect \BOthers {.}}{%
{Jiang}%
\ \protect \BOthers {.}}{%
{\protect \APACyear {2021}}%
}]{%
jiangetal21}
\APACinsertmetastar {%
jiangetal21}%
\begin{APACrefauthors}%
{Jiang}, B\BHBI W.%
, {Marziani}, P.%
, {Savi{\'c}}, {\DJ}.%
\ et al.\end{APACrefauthors}%
\unskip\
\newblock
\APACrefYearMonthDay{2021}{{\APACmonth{08}}}{},
\newblock
\unskip
\newblock
\APACjournalVolNumPages{\mnras}{}{}{}.
\newblock
\begin{APACrefDOI} \doi{10.1093/mnras/stab2273} \end{APACrefDOI}
\PrintBackRefs{\CurrentBib}

\bibitem [\protect \citeauthoryear {%
{Kaspi}%
\ \protect \BOthers {.}}{%
{Kaspi}%
\ \protect \BOthers {.}}{%
{\protect \APACyear {2000}}%
}]{%
kaspietal00}
\APACinsertmetastar {%
kaspietal00}%
\begin{APACrefauthors}%
{Kaspi}, S.%
, {Smith}, P\BPBI S.%
, {Netzer}, H.%
, {Maoz}, D.%
, {Jannuzi}, B\BPBI T.%
\BCBL {}\ \BBA {} {Giveon}, U.%
\end{APACrefauthors}%
\unskip\
\newblock
\APACrefYearMonthDay{2000}{{\APACmonth{04}}}{},
\newblock
\unskip
\newblock
\APACjournalVolNumPages{\apj}{533}{}{631-649}.
\newblock
\begin{APACrefDOI} \doi{10.1086/308704} \end{APACrefDOI}
\PrintBackRefs{\CurrentBib}

\bibitem [\protect \citeauthoryear {%
{Komossa}%
\ \protect \BOthers {.}}{%
{Komossa}%
\ \protect \BOthers {.}}{%
{\protect \APACyear {2006}}%
}]{%
komossaetal06}
\APACinsertmetastar {%
komossaetal06}%
\begin{APACrefauthors}%
{Komossa}, S.%
, {Voges}, W.%
, {Xu}, D.%
\ et al.\end{APACrefauthors}%
\unskip\
\newblock
\APACrefYearMonthDay{2006}{{\APACmonth{08}}}{},
\newblock
\unskip
\newblock
\APACjournalVolNumPages{\aj}{132}{}{531-545}.
\newblock
\begin{APACrefDOI} \doi{10.1086/505043} \end{APACrefDOI}
\PrintBackRefs{\CurrentBib}

\bibitem [\protect \citeauthoryear {%
{Laor}%
, {Fiore}%
, {Elvis}%
, {Wilkes}%
\BCBL {}\ \BBA {} {McDowell}%
}{%
{Laor}%
\ \protect \BOthers {.}}{%
{\protect \APACyear {1997}}%
}]{%
laoretal97b}
\APACinsertmetastar {%
laoretal97b}%
\begin{APACrefauthors}%
{Laor}, A.%
, {Fiore}, F.%
, {Elvis}, M.%
, {Wilkes}, B\BPBI J.%
\BCBL {}\ \BBA {} {McDowell}, J\BPBI C.%
\end{APACrefauthors}%
\unskip\
\newblock
\APACrefYearMonthDay{1997}{{\APACmonth{03}}}{},
\newblock
\unskip
\newblock
\APACjournalVolNumPages{ApJ}{477}{}{93-+}.
\newblock
\begin{APACrefDOI} \doi{10.1086/303696} \end{APACrefDOI}
\PrintBackRefs{\CurrentBib}

\bibitem [\protect \citeauthoryear {%
{Marin}%
}{%
{Marin}%
}{%
{\protect \APACyear {2016}}%
}]{%
marinetal16}
\APACinsertmetastar {%
marinetal16}%
\begin{APACrefauthors}%
{Marin}, F.%
\end{APACrefauthors}%
\unskip\
\newblock
\APACrefYearMonthDay{2016}{{\APACmonth{08}}}{},
\newblock
\unskip
\newblock
\APACjournalVolNumPages{\mnras}{460}{4}{3679-3705}.
\newblock
\begin{APACrefDOI} \doi{10.1093/mnras/stw1131} \end{APACrefDOI}
\PrintBackRefs{\CurrentBib}

\bibitem [\protect \citeauthoryear {%
{Marinello}%
, {Rodriguez-Ardila}%
, {Garcia-Rissmann}%
, {Sigut}%
\BCBL {}\ \BBA {} {Pradhan}%
}{%
{Marinello}%
\ \protect \BOthers {.}}{%
{\protect \APACyear {2016}}%
}]{%
marinelloetal16}
\APACinsertmetastar {%
marinelloetal16}%
\begin{APACrefauthors}%
{Marinello}, A\BPBI O\BPBI M.%
, {Rodriguez-Ardila}, A.%
, {Garcia-Rissmann}, A.%
, {Sigut}, T\BPBI A\BPBI A.%
\BCBL {}\ \BBA {} {Pradhan}, A\BPBI K.%
\end{APACrefauthors}%
\unskip\
\newblock
\APACrefYearMonthDay{2016}{{\APACmonth{02}}}{},
\newblock
\unskip
\newblock
\APACjournalVolNumPages{ApJ}{820}{2}{116}.
\PrintBackRefs{\CurrentBib}

\bibitem [\protect \citeauthoryear {%
{Mart{\'\i}nez-Aldama}%
\ \protect \BOthers {.}}{%
{Mart{\'\i}nez-Aldama}%
\ \protect \BOthers {.}}{%
{\protect \APACyear {2019}}%
}]{%
martinez-aldamaetal20}
\APACinsertmetastar {%
martinez-aldamaetal20}%
\begin{APACrefauthors}%
{Mart{\'\i}nez-Aldama}, M\BPBI L.%
, {Czerny}, B.%
, {Kawka}, D.%
, {Karas}, V.%
, {Panda}, S.%
, {Zaja{\v{c}}ek}, M.%
\BCBL {}\ \BBA {} {{\.Z}ycki}, P\BPBI T.%
\end{APACrefauthors}%
\unskip\
\newblock
\APACrefYearMonthDay{2019}{{\APACmonth{10}}}{},
\newblock
\unskip
\newblock
\APACjournalVolNumPages{\apj}{883}{2}{170}.
\newblock
\begin{APACrefDOI} \doi{10.3847/1538-4357/ab3728} \end{APACrefDOI}
\PrintBackRefs{\CurrentBib}

\bibitem [\protect \citeauthoryear {%
{Mart{\'\i}nez-Aldama}%
\ \protect \BOthers {.}}{%
{Mart{\'\i}nez-Aldama}%
\ \protect \BOthers {.}}{%
{\protect \APACyear {2018}}%
}]{%
martinez-aldamaetal18a}
\APACinsertmetastar {%
martinez-aldamaetal18a}%
\begin{APACrefauthors}%
{Mart{\'\i}nez-Aldama}, M\BPBI L.%
, {del Olmo}, A.%
, {Marziani}, P.%
\ et al.\end{APACrefauthors}%
\unskip\
\newblock
\APACrefYearMonthDay{2018}{{\APACmonth{11}}}{},
\newblock
\unskip
\newblock
\APACjournalVolNumPages{\aap}{618}{}{A179}.
\newblock
\begin{APACrefDOI} \doi{10.1051/0004-6361/201833541} \end{APACrefDOI}
\PrintBackRefs{\CurrentBib}

\bibitem [\protect \citeauthoryear {%
{Marziani}%
, {Bon}%
\BCBL {}\ \protect \BOthers {.}}{%
{Marziani}%
, {Bon}%
\BCBL {}\ \protect \BOthers {.}}{%
{\protect \APACyear {2020}}%
}]{%
marzianietal20a}
\APACinsertmetastar {%
marzianietal20a}%
\begin{APACrefauthors}%
{Marziani}, P.%
, {Bon}, E.%
, {Bon}, N.%
\ et al.\end{APACrefauthors}%
\unskip\
\newblock
\APACrefYearMonthDay{2020}{{\APACmonth{01}}}{},
\newblock
\unskip
\newblock
\APACjournalVolNumPages{Contributions of the Astronomical Observatory Skalnate
  Pleso}{50}{1}{244-256}.
\newblock
\begin{APACrefDOI} \doi{10.31577/caosp.2020.50.1.244} \end{APACrefDOI}
\PrintBackRefs{\CurrentBib}

\bibitem [\protect \citeauthoryear {%
{Marziani}%
, {del Olmo}%
\BCBL {}\ \protect \BOthers {.}}{%
{Marziani}%
, {del Olmo}%
\BCBL {}\ \protect \BOthers {.}}{%
{\protect \APACyear {2018}}%
}]{%
marzianietal18a}
\APACinsertmetastar {%
marzianietal18a}%
\begin{APACrefauthors}%
{Marziani}, P.%
, {del Olmo}, A.%
, {D'Onofrio}, M.%
\ et al.\end{APACrefauthors}%
\unskip\
\newblock
\APACrefYearMonthDay{2018}{{\APACmonth{04}}}{},
\newblock
{\BBOQ}\APACrefatitle {{Narrow-line Seyfert 1s: what is wrong in a name?}}
  {{Narrow-line Seyfert 1s: what is wrong in a name?}}{\BBCQ}
\newblock
\BIn{} \APACrefbtitle {Revisiting narrow-line Seyfert 1 galaxies and their
  place in the Universe. 9-13 April 2018. Padova Botanical Garden, Italy.
  Online at <A
  href=``https://pos.sissa.it/cgi-bin/reader/conf.cgi?confid=328''>https://pos.sissa.it/cgi-bin/reader/conf.cgi?confid=328</A>,
  id.2} {Revisiting narrow-line Seyfert 1 galaxies and their place in the
  Universe. 9-13 April 2018. Padova Botanical Garden, Italy. Online at <A
  href=``https://pos.sissa.it/cgi-bin/reader/conf.cgi?confid=328''>https://pos.sissa.it/cgi-bin/reader/conf.cgi?confid=328</A>,
  id.2}\ \BVOL\ PoS(NLS1-2018), \BPG~002.
\newblock
\APACaddressPublisher{}{SISSA/ISAS}.
\PrintBackRefs{\CurrentBib}

\bibitem [\protect \citeauthoryear {%
{Marziani}%
, {del Olmo}%
, {Perea}%
, {D'Onofrio}%
\BCBL {}\ \BBA {} {Panda}%
}{%
{Marziani}%
, {del Olmo}%
\BCBL {}\ \protect \BOthers {.}}{%
{\protect \APACyear {2020}}%
}]{%
marzianietal20}
\APACinsertmetastar {%
marzianietal20}%
\begin{APACrefauthors}%
{Marziani}, P.%
, {del Olmo}, A.%
, {Perea}, J.%
, {D'Onofrio}, M.%
\BCBL {}\ \BBA {} {Panda}, S.%
\end{APACrefauthors}%
\unskip\
\newblock
\APACrefYearMonthDay{2020}{{\APACmonth{12}}}{},
\newblock
\unskip
\newblock
\APACjournalVolNumPages{Atoms}{8}{4}{94}.
\newblock
\begin{APACrefDOI} \doi{10.3390/atoms8040094} \end{APACrefDOI}
\PrintBackRefs{\CurrentBib}

\bibitem [\protect \citeauthoryear {%
{Marziani}%
\ \protect \BOthers {.}}{%
{Marziani}%
\ \protect \BOthers {.}}{%
{\protect \APACyear {2021}}%
}]{%
marzianietal21}
\APACinsertmetastar {%
marzianietal21}%
\begin{APACrefauthors}%
{Marziani}, P.%
, {Dultzin}, D.%
, {del Olmo}, A.%
\ et al.\end{APACrefauthors}%
\unskip\
\newblock
\APACrefYearMonthDay{2021}{{\APACmonth{01}}}{},
\newblock
\unskip
\newblock
\APACjournalVolNumPages{IAU Symposium}{356}{}{66-71}.
\newblock
\begin{APACrefDOI} \doi{10.1017/S1743921320002598} \end{APACrefDOI}
\PrintBackRefs{\CurrentBib}

\bibitem [\protect \citeauthoryear {%
{Marziani}%
, {Dultzin}%
\BCBL {}\ \protect \BOthers {.}}{%
{Marziani}%
, {Dultzin}%
\BCBL {}\ \protect \BOthers {.}}{%
{\protect \APACyear {2018}}%
}]{%
marzianietal18}
\APACinsertmetastar {%
marzianietal18}%
\begin{APACrefauthors}%
{Marziani}, P.%
, {Dultzin}, D.%
, {Sulentic}, J\BPBI W.%
\ et al.\end{APACrefauthors}%
\unskip\
\newblock
\APACrefYearMonthDay{2018}{{\APACmonth{03}}}{},
\newblock
\unskip
\newblock
\APACjournalVolNumPages{Frontiers in Astronomy and Space Sciences}{5}{}{6}.
\newblock
\begin{APACrefDOI} \doi{10.3389/fspas.2018.00006} \end{APACrefDOI}
\PrintBackRefs{\CurrentBib}

\bibitem [\protect \citeauthoryear {%
{Marziani}%
\ \BBA {} {Sulentic}%
}{%
{Marziani}%
\ \BBA {} {Sulentic}%
}{%
{\protect \APACyear {2014}}%
}]{%
marzianisulentic14}
\APACinsertmetastar {%
marzianisulentic14}%
\begin{APACrefauthors}%
{Marziani}, P.%
\BCBT {}\ \BBA {} {Sulentic}, J\BPBI W.%
\end{APACrefauthors}%
\unskip\
\newblock
\APACrefYearMonthDay{2014}{{\APACmonth{08}}}{},
\newblock
\unskip
\newblock
\APACjournalVolNumPages{\mnras}{442}{}{1211-1229}.
\newblock
\begin{APACrefDOI} \doi{10.1093/mnras/stu951} \end{APACrefDOI}
\PrintBackRefs{\CurrentBib}

\bibitem [\protect \citeauthoryear {%
{Marziani}%
, {Sulentic}%
, {Dultzin-Hacyan}%
, {Calvani}%
\BCBL {}\ \BBA {} {Moles}%
}{%
{Marziani}%
\ \protect \BOthers {.}}{%
{\protect \APACyear {1996}}%
}]{%
marzianietal96}
\APACinsertmetastar {%
marzianietal96}%
\begin{APACrefauthors}%
{Marziani}, P.%
, {Sulentic}, J\BPBI W.%
, {Dultzin-Hacyan}, D.%
, {Calvani}, M.%
\BCBL {}\ \BBA {} {Moles}, M.%
\end{APACrefauthors}%
\unskip\
\newblock
\APACrefYearMonthDay{1996}{{\APACmonth{05}}}{},
\newblock
\unskip
\newblock
\APACjournalVolNumPages{ApJS}{104}{}{37-+}.
\newblock
\begin{APACrefDOI} \doi{10.1086/192291} \end{APACrefDOI}
\PrintBackRefs{\CurrentBib}

\bibitem [\protect \citeauthoryear {%
{Marziani}%
\ \protect \BOthers {.}}{%
{Marziani}%
\ \protect \BOthers {.}}{%
{\protect \APACyear {2014}}%
}]{%
marzianietal14}
\APACinsertmetastar {%
marzianietal14}%
\begin{APACrefauthors}%
{Marziani}, P.%
, {Sulentic}, J\BPBI W.%
, {Negrete}, C\BPBI A.%
, {Dultzin}, D.%
, {D'Onofrio}, M.%
, {Del Olmo}, A.%
\BCBL {}\ \BBA {} {Martinez-Aldama}, M\BPBI L.%
\end{APACrefauthors}%
\unskip\
\newblock
\APACrefYearMonthDay{2014}{{\APACmonth{10}}}{},
\newblock
\unskip
\newblock
\APACjournalVolNumPages{The Astronomical Review}{9}{}{6-25}.
\PrintBackRefs{\CurrentBib}

\bibitem [\protect \citeauthoryear {%
{Marziani}%
\ \protect \BOthers {.}}{%
{Marziani}%
\ \protect \BOthers {.}}{%
{\protect \APACyear {2010}}%
}]{%
marzianietal10}
\APACinsertmetastar {%
marzianietal10}%
\begin{APACrefauthors}%
{Marziani}, P.%
, {Sulentic}, J\BPBI W.%
, {Negrete}, C\BPBI A.%
, {Dultzin}, D.%
, {Zamfir}, S.%
\BCBL {}\ \BBA {} {Bachev}, R.%
\end{APACrefauthors}%
\unskip\
\newblock
\APACrefYearMonthDay{2010}{{\APACmonth{12}}}{},
\newblock
\unskip
\newblock
\APACjournalVolNumPages{\mnras}{409}{}{1033-1048}.
\newblock
\begin{APACrefDOI} \doi{10.1111/j.1365-2966.2010.17357.x} \end{APACrefDOI}
\PrintBackRefs{\CurrentBib}

\bibitem [\protect \citeauthoryear {%
{Marziani}%
, {Sulentic}%
, {Plauchu-Frayn}%
\BCBL {}\ \BBA {} {del Olmo}%
}{%
{Marziani}%
\ \protect \BOthers {.}}{%
{\protect \APACyear {2013}}%
}]{%
marzianietal13a}
\APACinsertmetastar {%
marzianietal13a}%
\begin{APACrefauthors}%
{Marziani}, P.%
, {Sulentic}, J\BPBI W.%
, {Plauchu-Frayn}, I.%
\BCBL {}\ \BBA {} {del Olmo}, A.%
\end{APACrefauthors}%
\unskip\
\newblock
\APACrefYearMonthDay{2013}{{\APACmonth{05}}}{},
\newblock
\unskip
\newblock
\APACjournalVolNumPages{AAp}{555}{}{89, 16pp}.
\PrintBackRefs{\CurrentBib}

\bibitem [\protect \citeauthoryear {%
{Marziani}%
, {Sulentic}%
, {Zwitter}%
, {Dultzin-Hacyan}%
\BCBL {}\ \BBA {} {Calvani}%
}{%
{Marziani}%
\ \protect \BOthers {.}}{%
{\protect \APACyear {2001}}%
}]{%
marzianietal01}
\APACinsertmetastar {%
marzianietal01}%
\begin{APACrefauthors}%
{Marziani}, P.%
, {Sulentic}, J\BPBI W.%
, {Zwitter}, T.%
, {Dultzin-Hacyan}, D.%
\BCBL {}\ \BBA {} {Calvani}, M.%
\end{APACrefauthors}%
\unskip\
\newblock
\APACrefYearMonthDay{2001}{{\APACmonth{09}}}{},
\newblock
\unskip
\newblock
\APACjournalVolNumPages{ApJ}{558}{}{553-560}.
\newblock
\begin{APACrefDOI} \doi{10.1086/322286} \end{APACrefDOI}
\PrintBackRefs{\CurrentBib}

\bibitem [\protect \citeauthoryear {%
{Marziani}%
, {Zamanov}%
, {Sulentic}%
\BCBL {}\ \BBA {} {Calvani}%
}{%
{Marziani}%
\ \protect \BOthers {.}}{%
{\protect \APACyear {2003}}%
}]{%
marzianietal03b}
\APACinsertmetastar {%
marzianietal03b}%
\begin{APACrefauthors}%
{Marziani}, P.%
, {Zamanov}, R\BPBI K.%
, {Sulentic}, J\BPBI W.%
\BCBL {}\ \BBA {} {Calvani}, M.%
\end{APACrefauthors}%
\unskip\
\newblock
\APACrefYearMonthDay{2003}{{\APACmonth{11}}}{},
\newblock
\unskip
\newblock
\APACjournalVolNumPages{MNRAS}{345}{}{1133-1144}.
\newblock
\begin{APACrefDOI} \doi{10.1046/j.1365-2966.2003.07033.x} \end{APACrefDOI}
\PrintBackRefs{\CurrentBib}

\bibitem [\protect \citeauthoryear {%
{Mineshige}%
, {Kawaguchi}%
, {Takeuchi}%
\BCBL {}\ \BBA {} {Hayashida}%
}{%
{Mineshige}%
\ \protect \BOthers {.}}{%
{\protect \APACyear {2000}}%
}]{%
mineshigeetal00}
\APACinsertmetastar {%
mineshigeetal00}%
\begin{APACrefauthors}%
{Mineshige}, S.%
, {Kawaguchi}, T.%
, {Takeuchi}, M.%
\BCBL {}\ \BBA {} {Hayashida}, K.%
\end{APACrefauthors}%
\unskip\
\newblock
\APACrefYearMonthDay{2000}{{\APACmonth{06}}}{},
\newblock
\unskip
\newblock
\APACjournalVolNumPages{\pasj}{52}{}{499-508}.
\PrintBackRefs{\CurrentBib}

\bibitem [\protect \citeauthoryear {%
{Negrete}%
\ \protect \BOthers {.}}{%
{Negrete}%
\ \protect \BOthers {.}}{%
{\protect \APACyear {2018}}%
}]{%
negreteetal18}
\APACinsertmetastar {%
negreteetal18}%
\begin{APACrefauthors}%
{Negrete}, C\BPBI A.%
, {Dultzin}, D.%
, {Marziani}, P.%
\ et al.\end{APACrefauthors}%
\unskip\
\newblock
\APACrefYearMonthDay{2018}{{\APACmonth{12}}}{},
\newblock
\unskip
\newblock
\APACjournalVolNumPages{\aap}{620}{}{A118}.
\newblock
\begin{APACrefDOI} \doi{10.1051/0004-6361/201833285} \end{APACrefDOI}
\PrintBackRefs{\CurrentBib}

\bibitem [\protect \citeauthoryear {%
{Padovani}%
}{%
{Padovani}%
}{%
{\protect \APACyear {2017}}%
}]{%
padovani17}
\APACinsertmetastar {%
padovani17}%
\begin{APACrefauthors}%
{Padovani}, P.%
\end{APACrefauthors}%
\unskip\
\newblock
\APACrefYearMonthDay{2017}{{\APACmonth{11}}}{},
\newblock
\unskip
\newblock
\APACjournalVolNumPages{Frontiers in Astronomy and Space Sciences}{4}{}{35}.
\newblock
\begin{APACrefDOI} \doi{10.3389/fspas.2017.00035} \end{APACrefDOI}
\PrintBackRefs{\CurrentBib}

\bibitem [\protect \citeauthoryear {%
Panda%
\ \protect \BOthers {.}}{%
Panda%
\ \protect \BOthers {.}}{%
{\protect \APACyear {2018}}%
}]{%
pandaetal18}
\APACinsertmetastar {%
pandaetal18}%
\begin{APACrefauthors}%
Panda, S.%
, Czerny, B.%
, Adhikari, T\BPBI P.%
, Hryniewicz, K.%
, Wildy, C.%
, Kuraszkiewicz, J.%
\BCBL {}\ \BBA {} {\'{S}}niegowska, M.%
\end{APACrefauthors}%
\unskip\
\newblock
\APACrefYearMonthDay{2018}{oct}{},
\newblock
\unskip
\newblock
\APACjournalVolNumPages{The Astrophysical Journal}{866}{2}{115}.
\newblock
\begin{APACrefURL} \url{https://doi.org/10.3847%2F1538-4357%2Faae209}
  \end{APACrefURL}
\newblock
\begin{APACrefDOI} \doi{10.3847/1538-4357/aae209} \end{APACrefDOI}
\PrintBackRefs{\CurrentBib}

\bibitem [\protect \citeauthoryear {%
{Panda}%
, {Marziani}%
\BCBL {}\ \BBA {} {Czerny}%
}{%
{Panda}%
\ \protect \BOthers {.}}{%
{\protect \APACyear {2019}}%
}]{%
pandaetal19}
\APACinsertmetastar {%
pandaetal19}%
\begin{APACrefauthors}%
{Panda}, S.%
, {Marziani}, P.%
\BCBL {}\ \BBA {} {Czerny}, B.%
\end{APACrefauthors}%
\unskip\
\newblock
\APACrefYearMonthDay{2019}{{\APACmonth{09}}}{},
\newblock
\unskip
\newblock
\APACjournalVolNumPages{\apj}{882}{2}{79}.
\newblock
\begin{APACrefDOI} \doi{10.3847/1538-4357/ab3292} \end{APACrefDOI}
\PrintBackRefs{\CurrentBib}

\bibitem [\protect \citeauthoryear {%
{Peruzzi}%
\ \protect \BOthers {.}}{%
{Peruzzi}%
\ \protect \BOthers {.}}{%
{\protect \APACyear {2021}}%
}]{%
peruzzietal21}
\APACinsertmetastar {%
peruzzietal21}%
\begin{APACrefauthors}%
{Peruzzi}, T.%
, {Pasquato}, M.%
, {Ciroi}, S.%
, {Berton}, M.%
, {Marziani}, P.%
\BCBL {}\ \BBA {} {Nardini}, E.%
\end{APACrefauthors}%
\unskip\
\newblock
\APACrefYearMonthDay{2021}{{\APACmonth{08}}}{},
\newblock
\unskip
\newblock
\APACjournalVolNumPages{\aap}{652}{}{A19}.
\newblock
\begin{APACrefDOI} \doi{10.1051/0004-6361/202038911} \end{APACrefDOI}
\PrintBackRefs{\CurrentBib}

\bibitem [\protect \citeauthoryear {%
{Popovi{\'c}}%
, {Mediavilla}%
, {Bon}%
\BCBL {}\ \BBA {} {Ili{\'c}}%
}{%
{Popovi{\'c}}%
\ \protect \BOthers {.}}{%
{\protect \APACyear {2004}}%
}]{%
popovicetal04}
\APACinsertmetastar {%
popovicetal04}%
\begin{APACrefauthors}%
{Popovi{\'c}}, L\BPBI {\v{C}}.%
, {Mediavilla}, E.%
, {Bon}, E.%
\BCBL {}\ \BBA {} {Ili{\'c}}, D.%
\end{APACrefauthors}%
\unskip\
\newblock
\APACrefYearMonthDay{2004}{{\APACmonth{09}}}{},
\newblock
\unskip
\newblock
\APACjournalVolNumPages{\aap}{423}{}{909-918}.
\newblock
\begin{APACrefDOI} \doi{10.1051/0004-6361:20034431} \end{APACrefDOI}
\PrintBackRefs{\CurrentBib}

\bibitem [\protect \citeauthoryear {%
{Popovic}%
, {Vince}%
, {Atanackovic-Vukmanovic}%
\BCBL {}\ \BBA {} {Kubicela}%
}{%
{Popovic}%
\ \protect \BOthers {.}}{%
{\protect \APACyear {1995}}%
}]{%
popovicetal95}
\APACinsertmetastar {%
popovicetal95}%
\begin{APACrefauthors}%
{Popovic}, L\BPBI C.%
, {Vince}, I.%
, {Atanackovic-Vukmanovic}, O.%
\BCBL {}\ \BBA {} {Kubicela}, A.%
\end{APACrefauthors}%
\unskip\
\newblock
\APACrefYearMonthDay{1995}{{\APACmonth{01}}}{},
\newblock
\unskip
\newblock
\APACjournalVolNumPages{\aap}{293}{}{309-314}.
\PrintBackRefs{\CurrentBib}

\bibitem [\protect \citeauthoryear {%
{Punsly}%
}{%
{Punsly}%
}{%
{\protect \APACyear {2010}}%
}]{%
punsly10}
\APACinsertmetastar {%
punsly10}%
\begin{APACrefauthors}%
{Punsly}, B.%
\end{APACrefauthors}%
\unskip\
\newblock
\APACrefYearMonthDay{2010}{{\APACmonth{04}}}{},
\newblock
\unskip
\newblock
\APACjournalVolNumPages{\apj}{713}{}{232-238}.
\newblock
\begin{APACrefDOI} \doi{10.1088/0004-637X/713/1/232} \end{APACrefDOI}
\PrintBackRefs{\CurrentBib}

\bibitem [\protect \citeauthoryear {%
{Punsly}%
, {Marziani}%
, {Bennert}%
, {Nagai}%
\BCBL {}\ \BBA {} {Gurwell}%
}{%
{Punsly}%
, {Marziani}%
\BCBL {}\ \protect \BOthers {.}}{%
{\protect \APACyear {2018}}%
}]{%
punslyetal18a}
\APACinsertmetastar {%
punslyetal18a}%
\begin{APACrefauthors}%
{Punsly}, B.%
, {Marziani}, P.%
, {Bennert}, V\BPBI N.%
, {Nagai}, H.%
\BCBL {}\ \BBA {} {Gurwell}, M\BPBI A.%
\end{APACrefauthors}%
\unskip\
\newblock
\APACrefYearMonthDay{2018}{{\APACmonth{12}}}{},
\newblock
\unskip
\newblock
\APACjournalVolNumPages{\apj}{869}{}{143}.
\newblock
\begin{APACrefDOI} \doi{10.3847/1538-4357/aaec75} \end{APACrefDOI}
\PrintBackRefs{\CurrentBib}

\bibitem [\protect \citeauthoryear {%
{Punsly}%
, {Marziani}%
, {Berton}%
\BCBL {}\ \BBA {} {Kharb}%
}{%
{Punsly}%
\ \protect \BOthers {.}}{%
{\protect \APACyear {2020}}%
}]{%
punslyetal20}
\APACinsertmetastar {%
punslyetal20}%
\begin{APACrefauthors}%
{Punsly}, B.%
, {Marziani}, P.%
, {Berton}, M.%
\BCBL {}\ \BBA {} {Kharb}, P.%
\end{APACrefauthors}%
\unskip\
\newblock
\APACrefYearMonthDay{2020}{{\APACmonth{11}}}{},
\newblock
\unskip
\newblock
\APACjournalVolNumPages{\apj}{903}{1}{44}.
\newblock
\begin{APACrefDOI} \doi{10.3847/1538-4357/abb950} \end{APACrefDOI}
\PrintBackRefs{\CurrentBib}

\bibitem [\protect \citeauthoryear {%
{Punsly}%
, {Tramacere}%
, {Kharb}%
\BCBL {}\ \BBA {} {Marziani}%
}{%
{Punsly}%
, {Tramacere}%
\BCBL {}\ \protect \BOthers {.}}{%
{\protect \APACyear {2018}}%
}]{%
punslyetal18}
\APACinsertmetastar {%
punslyetal18}%
\begin{APACrefauthors}%
{Punsly}, B.%
, {Tramacere}, A.%
, {Kharb}, P.%
\BCBL {}\ \BBA {} {Marziani}, P.%
\end{APACrefauthors}%
\unskip\
\newblock
\APACrefYearMonthDay{2018}{{\APACmonth{12}}}{},
\newblock
\unskip
\newblock
\APACjournalVolNumPages{\apj}{869}{2}{174}.
\newblock
\begin{APACrefDOI} \doi{10.3847/1538-4357/aaefe7} \end{APACrefDOI}
\PrintBackRefs{\CurrentBib}

\bibitem [\protect \citeauthoryear {%
{Richards}%
\ \protect \BOthers {.}}{%
{Richards}%
\ \protect \BOthers {.}}{%
{\protect \APACyear {2011}}%
}]{%
richardsetal11}
\APACinsertmetastar {%
richardsetal11}%
\begin{APACrefauthors}%
{Richards}, G\BPBI T.%
, {Kruczek}, N\BPBI E.%
, {Gallagher}, S\BPBI C.%
\ et al.\end{APACrefauthors}%
\unskip\
\newblock
\APACrefYearMonthDay{2011}{{\APACmonth{05}}}{},
\newblock
\unskip
\newblock
\APACjournalVolNumPages{\aj}{141}{}{167-+}.
\newblock
\begin{APACrefDOI} \doi{10.1088/0004-6256/141/5/167} \end{APACrefDOI}
\PrintBackRefs{\CurrentBib}

\bibitem [\protect \citeauthoryear {%
{Shen}%
\ \BBA {} {Ho}%
}{%
{Shen}%
\ \BBA {} {Ho}%
}{%
{\protect \APACyear {2014}}%
}]{%
shenho14}
\APACinsertmetastar {%
shenho14}%
\begin{APACrefauthors}%
{Shen}, Y.%
\BCBT {}\ \BBA {} {Ho}, L\BPBI C.%
\end{APACrefauthors}%
\unskip\
\newblock
\APACrefYearMonthDay{2014}{{\APACmonth{09}}}{},
\newblock
\unskip
\newblock
\APACjournalVolNumPages{\nat}{513}{}{210-213}.
\newblock
\begin{APACrefDOI} \doi{10.1038/nature13712} \end{APACrefDOI}
\PrintBackRefs{\CurrentBib}

\bibitem [\protect \citeauthoryear {%
{{\'S}niegowska}%
\ \protect \BOthers {.}}{%
{{\'S}niegowska}%
\ \protect \BOthers {.}}{%
{\protect \APACyear {2021}}%
}]{%
sniegowskaetal21}
\APACinsertmetastar {%
sniegowskaetal21}%
\begin{APACrefauthors}%
{{\'S}niegowska}, M.%
, {Marziani}, P.%
, {Czerny}, B.%
, {Panda}, S.%
, {Mart{\'\i}nez-Aldama}, M\BPBI L.%
, {del Olmo}, A.%
\BCBL {}\ \BBA {} {D'Onofrio}, M.%
\end{APACrefauthors}%
\unskip\
\newblock
\APACrefYearMonthDay{2021}{{\APACmonth{04}}}{},
\newblock
\unskip
\newblock
\APACjournalVolNumPages{\apj}{910}{2}{115}.
\newblock
\begin{APACrefDOI} \doi{10.3847/1538-4357/abe1c8} \end{APACrefDOI}
\PrintBackRefs{\CurrentBib}

\bibitem [\protect \citeauthoryear {%
{Strateva}%
\ \protect \BOthers {.}}{%
{Strateva}%
\ \protect \BOthers {.}}{%
{\protect \APACyear {2003}}%
}]{%
stratevaetal03}
\APACinsertmetastar {%
stratevaetal03}%
\begin{APACrefauthors}%
{Strateva}, I\BPBI V.%
, {Strauss}, M\BPBI A.%
, {Hao}, L.%
\ et al.\end{APACrefauthors}%
\unskip\
\newblock
\APACrefYearMonthDay{2003}{{\APACmonth{10}}}{},
\newblock
\unskip
\newblock
\APACjournalVolNumPages{AJ}{126}{}{1720-1749}.
\newblock
\begin{APACrefDOI} \doi{10.1086/378367} \end{APACrefDOI}
\PrintBackRefs{\CurrentBib}

\bibitem [\protect \citeauthoryear {%
{Sulentic}%
, {Bachev}%
, {Marziani}%
, {Negrete}%
\BCBL {}\ \BBA {} {Dultzin}%
}{%
{Sulentic}%
\ \protect \BOthers {.}}{%
{\protect \APACyear {2007}}%
}]{%
sulenticetal07}
\APACinsertmetastar {%
sulenticetal07}%
\begin{APACrefauthors}%
{Sulentic}, J\BPBI W.%
, {Bachev}, R.%
, {Marziani}, P.%
, {Negrete}, C\BPBI A.%
\BCBL {}\ \BBA {} {Dultzin}, D.%
\end{APACrefauthors}%
\unskip\
\newblock
\APACrefYearMonthDay{2007}{{\APACmonth{09}}}{},
\newblock
\unskip
\newblock
\APACjournalVolNumPages{\apj}{666}{2}{757-777}.
\newblock
\begin{APACrefDOI} \doi{10.1086/519916} \end{APACrefDOI}
\PrintBackRefs{\CurrentBib}

\bibitem [\protect \citeauthoryear {%
{Sulentic}%
, {Calvani}%
, {Marziani}%
\BCBL {}\ \BBA {} {Zheng}%
}{%
{Sulentic}%
\ \protect \BOthers {.}}{%
{\protect \APACyear {1990}}%
}]{%
sulenticetal90}
\APACinsertmetastar {%
sulenticetal90}%
\begin{APACrefauthors}%
{Sulentic}, J\BPBI W.%
, {Calvani}, M.%
, {Marziani}, P.%
\BCBL {}\ \BBA {} {Zheng}, W.%
\end{APACrefauthors}%
\unskip\
\newblock
\APACrefYearMonthDay{1990}{{\APACmonth{05}}}{},
\newblock
\unskip
\newblock
\APACjournalVolNumPages{\apjl}{355}{}{L15}.
\newblock
\begin{APACrefDOI} \doi{10.1086/185727} \end{APACrefDOI}
\PrintBackRefs{\CurrentBib}

\bibitem [\protect \citeauthoryear {%
{Sulentic}%
\ \protect \BOthers {.}}{%
{Sulentic}%
\ \protect \BOthers {.}}{%
{\protect \APACyear {2015}}%
}]{%
sulenticetal15}
\APACinsertmetastar {%
sulenticetal15}%
\begin{APACrefauthors}%
{Sulentic}, J\BPBI W.%
, {Mart{\'{\i}}nez-Carballo}, M\BPBI A.%
, {Marziani}, P.%
, {del Olmo}, A.%
, {Stirpe}, G\BPBI M.%
, {Zamfir}, S.%
\BCBL {}\ \BBA {} {Plauchu-Frayn}, I.%
\end{APACrefauthors}%
\unskip\
\newblock
\APACrefYearMonthDay{2015}{{\APACmonth{06}}}{},
\newblock
\unskip
\newblock
\APACjournalVolNumPages{\mnras}{450}{}{1916-1925}.
\newblock
\begin{APACrefDOI} \doi{10.1093/mnras/stv710} \end{APACrefDOI}
\PrintBackRefs{\CurrentBib}

\bibitem [\protect \citeauthoryear {%
{Sulentic}%
, {Marziani}%
\BCBL {}\ \BBA {} {Dultzin-Hacyan}%
}{%
{Sulentic}%
, {Marziani}%
\BCBL {}\ \BBA {} {Dultzin-Hacyan}%
}{%
{\protect \APACyear {2000}}%
}]{%
sulenticetal00a}
\APACinsertmetastar {%
sulenticetal00a}%
\begin{APACrefauthors}%
{Sulentic}, J\BPBI W.%
, {Marziani}, P.%
\BCBL {}\ \BBA {} {Dultzin-Hacyan}, D.%
\end{APACrefauthors}%
\unskip\
\newblock
\APACrefYearMonthDay{2000}{}{},
\newblock
\unskip
\newblock
\APACjournalVolNumPages{ARA\&A}{38}{}{521-571}.
\newblock
\begin{APACrefDOI} \doi{10.1146/annurev.astro.38.1.521} \end{APACrefDOI}
\PrintBackRefs{\CurrentBib}

\bibitem [\protect \citeauthoryear {%
{Sulentic}%
\ \protect \BOthers {.}}{%
{Sulentic}%
\ \protect \BOthers {.}}{%
{\protect \APACyear {2002}}%
}]{%
sulenticetal02}
\APACinsertmetastar {%
sulenticetal02}%
\begin{APACrefauthors}%
{Sulentic}, J\BPBI W.%
, {Marziani}, P.%
, {Zamanov}, R.%
, {Bachev}, R.%
, {Calvani}, M.%
\BCBL {}\ \BBA {} {Dultzin-Hacyan}, D.%
\end{APACrefauthors}%
\unskip\
\newblock
\APACrefYearMonthDay{2002}{{\APACmonth{02}}}{},
\newblock
\unskip
\newblock
\APACjournalVolNumPages{ApJL}{566}{}{L71-L75}.
\newblock
\begin{APACrefDOI} \doi{10.1086/339594} \end{APACrefDOI}
\PrintBackRefs{\CurrentBib}

\bibitem [\protect \citeauthoryear {%
{Sulentic}%
, {Marziani}%
, {Zwitter}%
, {Dultzin-Hacyan}%
\BCBL {}\ \BBA {} {Calvani}%
}{%
{Sulentic}%
, {Marziani}%
, {Zwitter}%
\BCBL {}\ \protect \BOthers {.}}{%
{\protect \APACyear {2000}}%
}]{%
sulenticetal00b}
\APACinsertmetastar {%
sulenticetal00b}%
\begin{APACrefauthors}%
{Sulentic}, J\BPBI W.%
, {Marziani}, P.%
, {Zwitter}, T.%
, {Dultzin-Hacyan}, D.%
\BCBL {}\ \BBA {} {Calvani}, M.%
\end{APACrefauthors}%
\unskip\
\newblock
\APACrefYearMonthDay{2000}{{\APACmonth{12}}}{},
\newblock
\unskip
\newblock
\APACjournalVolNumPages{ApJL}{545}{}{L15-L18}.
\newblock
\begin{APACrefDOI} \doi{10.1086/317330} \end{APACrefDOI}
\PrintBackRefs{\CurrentBib}

\bibitem [\protect \citeauthoryear {%
{Sulentic}%
\ \protect \BOthers {.}}{%
{Sulentic}%
\ \protect \BOthers {.}}{%
{\protect \APACyear {2003}}%
}]{%
sulenticetal03}
\APACinsertmetastar {%
sulenticetal03}%
\begin{APACrefauthors}%
{Sulentic}, J\BPBI W.%
, {Zamfir}, S.%
, {Marziani}, P.%
, {Bachev}, R.%
, {Calvani}, M.%
\BCBL {}\ \BBA {} {Dultzin-Hacyan}, D.%
\end{APACrefauthors}%
\unskip\
\newblock
\APACrefYearMonthDay{2003}{{\APACmonth{11}}}{},
\newblock
\unskip
\newblock
\APACjournalVolNumPages{ApJL}{597}{}{L17-L20}.
\newblock
\begin{APACrefDOI} \doi{10.1086/379754} \end{APACrefDOI}
\PrintBackRefs{\CurrentBib}

\bibitem [\protect \citeauthoryear {%
{Sun}%
\ \BBA {} {Shen}%
}{%
{Sun}%
\ \BBA {} {Shen}%
}{%
{\protect \APACyear {2015}}%
}]{%
sunshen15}
\APACinsertmetastar {%
sunshen15}%
\begin{APACrefauthors}%
{Sun}, J.%
\BCBT {}\ \BBA {} {Shen}, Y.%
\end{APACrefauthors}%
\unskip\
\newblock
\APACrefYearMonthDay{2015}{{\APACmonth{05}}}{},
\newblock
\unskip
\newblock
\APACjournalVolNumPages{\apjl}{804}{}{L15}.
\newblock
\begin{APACrefDOI} \doi{10.1088/2041-8205/804/1/L15} \end{APACrefDOI}
\PrintBackRefs{\CurrentBib}

\bibitem [\protect \citeauthoryear {%
{Temple}%
, {Ferland}%
, {Rankine}%
, {Chatzikos}%
\BCBL {}\ \BBA {} {Hewett}%
}{%
{Temple}%
\ \protect \BOthers {.}}{%
{\protect \APACyear {2021}}%
}]{%
templeetal21}
\APACinsertmetastar {%
templeetal21}%
\begin{APACrefauthors}%
{Temple}, M\BPBI J.%
, {Ferland}, G\BPBI J.%
, {Rankine}, A\BPBI L.%
, {Chatzikos}, M.%
\BCBL {}\ \BBA {} {Hewett}, P\BPBI C.%
\end{APACrefauthors}%
\unskip\
\newblock
\APACrefYearMonthDay{2021}{{\APACmonth{08}}}{},
\newblock
\unskip
\newblock
\APACjournalVolNumPages{\mnras}{505}{3}{3247-3259}.
\newblock
\begin{APACrefDOI} \doi{10.1093/mnras/stab1610} \end{APACrefDOI}
\PrintBackRefs{\CurrentBib}

\bibitem [\protect \citeauthoryear {%
{Urry}%
\ \BBA {} {Padovani}%
}{%
{Urry}%
\ \BBA {} {Padovani}%
}{%
{\protect \APACyear {1995}}%
}]{%
urrypadovani95}
\APACinsertmetastar {%
urrypadovani95}%
\begin{APACrefauthors}%
{Urry}, C\BPBI M.%
\BCBT {}\ \BBA {} {Padovani}, P.%
\end{APACrefauthors}%
\unskip\
\newblock
\APACrefYearMonthDay{1995}{{\APACmonth{09}}}{},
\newblock
\unskip
\newblock
\APACjournalVolNumPages{PASP}{107}{}{803}.
\newblock
\begin{APACrefDOI} \doi{10.1086/133630} \end{APACrefDOI}
\PrintBackRefs{\CurrentBib}

\bibitem [\protect \citeauthoryear {%
{Wang}%
\ \protect \BOthers {.}}{%
{Wang}%
\ \protect \BOthers {.}}{%
{\protect \APACyear {2017}}%
}]{%
wangetal17}
\APACinsertmetastar {%
wangetal17}%
\begin{APACrefauthors}%
{Wang}, J\BHBI M.%
, {Du}, P.%
, {Brotherton}, M\BPBI S.%
\ et al.\end{APACrefauthors}%
\unskip\
\newblock
\APACrefYearMonthDay{2017}{{\APACmonth{10}}}{},
\newblock
\unskip
\newblock
\APACjournalVolNumPages{Nature Astronomy}{1}{}{775-783}.
\newblock
\begin{APACrefDOI} \doi{10.1038/s41550-017-0264-4} \end{APACrefDOI}
\PrintBackRefs{\CurrentBib}

\bibitem [\protect \citeauthoryear {%
{Wang}%
\ \protect \BOthers {.}}{%
{Wang}%
\ \protect \BOthers {.}}{%
{\protect \APACyear {2014}}%
}]{%
wangetal14b}
\APACinsertmetastar {%
wangetal14b}%
\begin{APACrefauthors}%
{Wang}, J\BHBI M.%
, {Du}, P.%
, {Hu}, C.%
\ et al.\end{APACrefauthors}%
\unskip\
\newblock
\APACrefYearMonthDay{2014}{{\APACmonth{10}}}{},
\newblock
\unskip
\newblock
\APACjournalVolNumPages{\apj}{793}{2}{108}.
\newblock
\begin{APACrefDOI} \doi{10.1088/0004-637X/793/2/108} \end{APACrefDOI}
\PrintBackRefs{\CurrentBib}

\bibitem [\protect \citeauthoryear {%
{Wang}%
, {Du}%
, {Valls-Gabaud}%
, {Hu}%
\BCBL {}\ \BBA {} {Netzer}%
}{%
{Wang}%
\ \protect \BOthers {.}}{%
{\protect \APACyear {2013}}%
}]{%
wangetal13}
\APACinsertmetastar {%
wangetal13}%
\begin{APACrefauthors}%
{Wang}, J\BHBI M.%
, {Du}, P.%
, {Valls-Gabaud}, D.%
, {Hu}, C.%
\BCBL {}\ \BBA {} {Netzer}, H.%
\end{APACrefauthors}%
\unskip\
\newblock
\APACrefYearMonthDay{2013}{{\APACmonth{02}}}{},
\newblock
\unskip
\newblock
\APACjournalVolNumPages{Physical Review Letters}{110}{8}{081301}.
\newblock
\begin{APACrefDOI} \doi{10.1103/PhysRevLett.110.081301} \end{APACrefDOI}
\PrintBackRefs{\CurrentBib}

\bibitem [\protect \citeauthoryear {%
{Wills}%
\ \BBA {} {Brotherton}%
}{%
{Wills}%
\ \BBA {} {Brotherton}%
}{%
{\protect \APACyear {1995}}%
}]{%
willsbrotherton95}
\APACinsertmetastar {%
willsbrotherton95}%
\begin{APACrefauthors}%
{Wills}, B\BPBI J.%
\BCBT {}\ \BBA {} {Brotherton}, M\BPBI S.%
\end{APACrefauthors}%
\unskip\
\newblock
\APACrefYearMonthDay{1995}{{\APACmonth{08}}}{},
\newblock
\unskip
\newblock
\APACjournalVolNumPages{\apjl}{448}{}{L81}.
\newblock
\begin{APACrefDOI} \doi{10.1086/309614} \end{APACrefDOI}
\PrintBackRefs{\CurrentBib}

\bibitem [\protect \citeauthoryear {%
{Wills}%
\ \BBA {} {Browne}%
}{%
{Wills}%
\ \BBA {} {Browne}%
}{%
{\protect \APACyear {1986}}%
}]{%
willsbrowne86}
\APACinsertmetastar {%
willsbrowne86}%
\begin{APACrefauthors}%
{Wills}, B\BPBI J.%
\BCBT {}\ \BBA {} {Browne}, I\BPBI W\BPBI A.%
\end{APACrefauthors}%
\unskip\
\newblock
\APACrefYearMonthDay{1986}{{\APACmonth{03}}}{},
\newblock
\unskip
\newblock
\APACjournalVolNumPages{\apj}{302}{}{56-63}.
\newblock
\begin{APACrefDOI} \doi{10.1086/163973} \end{APACrefDOI}
\PrintBackRefs{\CurrentBib}

\bibitem [\protect \citeauthoryear {%
{Wolf}%
\ \protect \BOthers {.}}{%
{Wolf}%
\ \protect \BOthers {.}}{%
{\protect \APACyear {2020}}%
}]{%
wolfetal20}
\APACinsertmetastar {%
wolfetal20}%
\begin{APACrefauthors}%
{Wolf}, J.%
, {Salvato}, M.%
, {Coffey}, D.%
\ et al.\end{APACrefauthors}%
\unskip\
\newblock
\APACrefYearMonthDay{2020}{{\APACmonth{03}}}{},
\newblock
\unskip
\newblock
\APACjournalVolNumPages{\mnras}{492}{3}{3580-3601}.
\newblock
\begin{APACrefDOI} \doi{10.1093/mnras/staa018} \end{APACrefDOI}
\PrintBackRefs{\CurrentBib}

\bibitem [\protect \citeauthoryear {%
{Zamfir}%
, {Sulentic}%
\BCBL {}\ \BBA {} {Marziani}%
}{%
{Zamfir}%
\ \protect \BOthers {.}}{%
{\protect \APACyear {2008}}%
}]{%
zamfiretal08}
\APACinsertmetastar {%
zamfiretal08}%
\begin{APACrefauthors}%
{Zamfir}, S.%
, {Sulentic}, J\BPBI W.%
\BCBL {}\ \BBA {} {Marziani}, P.%
\end{APACrefauthors}%
\unskip\
\newblock
\APACrefYearMonthDay{2008}{{\APACmonth{06}}}{},
\newblock
\unskip
\newblock
\APACjournalVolNumPages{MNRAS}{387}{}{856-870}.
\newblock
\begin{APACrefDOI} \doi{10.1111/j.1365-2966.2008.13290.x} \end{APACrefDOI}
\PrintBackRefs{\CurrentBib}

\bibitem [\protect \citeauthoryear {%
{Zamfir}%
, {Sulentic}%
, {Marziani}%
\BCBL {}\ \BBA {} {Dultzin}%
}{%
{Zamfir}%
\ \protect \BOthers {.}}{%
{\protect \APACyear {2010}}%
}]{%
zamfiretal10}
\APACinsertmetastar {%
zamfiretal10}%
\begin{APACrefauthors}%
{Zamfir}, S.%
, {Sulentic}, J\BPBI W.%
, {Marziani}, P.%
\BCBL {}\ \BBA {} {Dultzin}, D.%
\end{APACrefauthors}%
\unskip\
\newblock
\APACrefYearMonthDay{2010}{{\APACmonth{02}}}{},
\newblock
\unskip
\newblock
\APACjournalVolNumPages{\mnras}{403}{}{1759}.
\newblock
\begin{APACrefDOI} \doi{10.1111/j.1365-2966.2009.16236.x} \end{APACrefDOI}
\PrintBackRefs{\CurrentBib}

\bibitem [\protect \citeauthoryear {%
{Zheng}%
\ \BBA {} {Sulentic}%
}{%
{Zheng}%
\ \BBA {} {Sulentic}%
}{%
{\protect \APACyear {1990}}%
}]{%
zhengsulentic90}
\APACinsertmetastar {%
zhengsulentic90}%
\begin{APACrefauthors}%
{Zheng}, W.%
\BCBT {}\ \BBA {} {Sulentic}, J\BPBI W.%
\end{APACrefauthors}%
\unskip\
\newblock
\APACrefYearMonthDay{1990}{{\APACmonth{02}}}{},
\newblock
\unskip
\newblock
\APACjournalVolNumPages{\apj}{350}{}{512}.
\newblock
\begin{APACrefDOI} \doi{10.1086/168407} \end{APACrefDOI}
\PrintBackRefs{\CurrentBib}

\bibitem [\protect \citeauthoryear {%
{Zheng}%
, {Sulentic}%
\BCBL {}\ \BBA {} {Binette}%
}{%
{Zheng}%
\ \protect \BOthers {.}}{%
{\protect \APACyear {1990}}%
}]{%
zhengetal90}
\APACinsertmetastar {%
zhengetal90}%
\begin{APACrefauthors}%
{Zheng}, W.%
, {Sulentic}, J\BPBI W.%
\BCBL {}\ \BBA {} {Binette}, L.%
\end{APACrefauthors}%
\unskip\
\newblock
\APACrefYearMonthDay{1990}{{\APACmonth{12}}}{},
\newblock
\unskip
\newblock
\APACjournalVolNumPages{\apj}{365}{}{115-118}.
\newblock
\begin{APACrefDOI} \doi{10.1086/169462} \end{APACrefDOI}
\PrintBackRefs{\CurrentBib}

\end{thebibliography}

\section*{Author Biography}
%
\begin{biography}{\includegraphics[width=60pt,height=60pt]{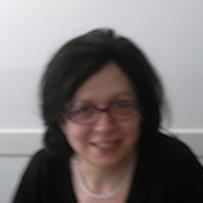}}{\textbf{Paola Marziani} is an astronomer with the National Institute of Astrophysics (INAF), based at the Padua Astronomical Observatory in Italy. }
\end{biography}

\end{document}